\documentclass[12pt,final]{article}

\IfFileExists{srcltx.sty}{\usepackage[active]{srcltx}}

\def\hybrid{\topmargin 0pt \oddsidemargin 0pt \headheight 0pt \headsep 0pt
       \textwidth 6.5in        
       \textheight 9in         
           \voffset=-0.3cm         
           \marginparwidth 0.0in \parskip 5pt plus 1pt \jot = 1.5ex}
     
     \hybrid \hfuzz=2pt

\makeatletter
\def\marginnote#1{}

\newcount\hour
\newcount\minute
\newtoks\amorpm
\hour=\time\divide\hour by60
\minute=\time{\multiply\hour by60 \global\advance\minute by-\hour}
\edef\standardtime{{\ifnum\hour<12 \global\amorpm={am}%
           \else\global\amorpm={pm}\advance\hour by-12 \fi
           \ifnum\hour=0 \hour=12 \fi
           \number\hour:\ifnum\minute<10 0\fi\number\minute\the\amorpm}}
\edef\militarytime{\number\hour:\ifnum\minute<10 0\fi\number\minute}

\def\draft{%
  \oddsidemargin -.5truein %
  \def\@oddfoot{\vbox{\scriptsize \hbox{\scriptsize File {\tt\jobname.tex}}
      \hbox{\scriptsize Draft $Revision: 1.14 $}} \hfil { -- \footnotesize
      \thepage\ -- } \hfil \vbox{\scriptsize \hbox to 0pt {\hss\scriptsize
        Time:
            \militarytime}\hbox to 0pt {\hss\scriptsize\today}}%
        \let\@evenfoot\@oddfoot \overfullrule 3pt
}%
}%

\def\numberbysection{\@addtoreset{equation}{section}
  \def\theequation{\thesection.\arabic{equation}}}

\makeatother

\relax 
\hybrid

\numberbysection

\usepackage{amsfonts,varioref,graphicx,xspace} %
\usepackage[intlimits]{amsmath} %
\usepackage{amssymb} %
\usepackage[notcite,notref]{showkeys}

\usepackage[colorlinks,bookmarks=true,bookmarksopenlevel=1]{hyperref}

\newcommand{\0}{{(0)}}
\newcommand{\1}{{(1)}}

\renewcommand{\vref}[1]{\ref{#1}~\vpageref{#1}\unskip\xspace}
\newcommand{\p}[1]{\partial_{#1}} %
\newcommand{\la}{\langle} %
\newcommand{\ra}{\rangle} %
\newcommand{\ket}[1]{\left|{#1}\right\rangle} %
\newcommand{\bra}[1]{\left\langle{#1}\right|} %
\newcommand{\Dsl}{\mathop{\lefteqn{D}{\,\mbox{\large /}}}\nolimits} 
\newcommand{\psl}{\mathop{\lefteqn{\partial}{\mbox{\large /}}}} 
\newcommand{\Asl}{\mathop{\lefteqn{A}{\,\mbox{\large /}}}} 
\newcommand{\zm}{{\mathrm{z.m}}} 
\newcommand{\cs}{{\textsc{cs}}} 
\newcommand{\eff}{{\textsc{eff}}} 
\newcommand{\df}{{\textsc{df}}} 
\newcommand{\z}{{\textsc{z}}} 
\newcommand{\ggz}{{\gamma\gamma\z}} 
\newcommand{\gzz}{{\gamma\z\z}} 
\newcommand{\zzz}{{\z\z\z}} 
\newcommand{\anom}{{\scriptstyle{\textsc{anom}}}} %
\newcommand{\Y}{{\scriptstyle{\textsc{y}}}} %
\newcommand{\Tr}{\mathop\mathrm{Tr}}  
\newcommand{\sign}{\mathop\mathrm{sign}} 
\newcommand{\const}{\mathrm{const}} %
\renewcommand{\div}{\mathop{\mathrm{div}}}
\newcommand{\e}{\,\mathbf{e}} 
\newcommand{\bE}{\text{\textbf{\slshape E}}} 
\newcommand{\bH}{\text{\textbf{\slshape H}}} 
\newcommand{\bF}{\text{\textbf{\slshape F}}} 
\newcommand{\bj}{\text{\textbf{\slshape j}}} 
\newcommand{\M}{M} 
\newcommand{\m}{\text{\textbf{\slshape m}}} 
\newcommand{\hs}{\hskip -.25em}
\def\CA{{\cal A}} %
\def\CE{{\cal E}}  %
\def\CF{{\cal F}}  %
\def\CJ{{\cal J}} %
\def\CO{{\cal O}}  %
\begin{document}
\title{Observational manifestations of anomaly inflow} %

\author{\large \sf Alexey Boyarsky\thanks{Ecole Polytechnique F\'ed\'erale de
    Lausanne, Institute of Theoretical Physics, FSB/ITP/LPPC, BSP 720,
    CH-1015, Lausanne, Switzerland},~~Oleg Ruchayskiy\thanks{Institut des
    Hautes Etudes Scientifiques, Bures-sur-Yvette, F-91440, France},~~Mikhail
  Shaposhnikov$^*$.} %
\date{} %
\maketitle
\begin{abstract}
  In theories with chiral couplings, one of the important consistency
  requirements is that of the cancellation of a gauge anomaly. In particular,
  this is one of the conditions imposed on the hypercharges in the Standard
  Model. However, anomaly cancellation condition of the Standard Model looks
  unnatural from the perspective of a theory with extra dimensions. Indeed, if
  our world were embedded into an odd-dimensional space, then the full theory
  would be automatically anomaly free. In this paper we discuss the physical
  consequences of anomaly non-cancellation for effective 4-dimensional field
  theory. We demonstrate that in such a theory parallel electric and magnetic
  fields get modified. In particular, this happens for any particle possessing
  both electric charge and magnetic moment. This effect, if observed, can
  serve as a \emph{low energy} signature of extra dimensions.  On the other
  hand, if such an effect is absent or is very small, then from the point of
  view of any theory with extra dimensions it is just another fine-tuning and
  should acquire theoretical explanation.
\end{abstract}


\section{Introduction}
\label{sec:intro}

In particle physics there is a number of well-established principles
(four-dimensional Lorentz invariance, renormalizability, gauge symmetries,
etc.), which are believed to underlie a consistent quantum field theory.  All
these principles are built into the Standard Model (SM) and its great
experimental success is a major justification for their validity.  However, in
continuing search for the consistent description of an emerging physics beyond
the Standard Model, one often tries to ease some of these criteria. It is then
important to make sure that such modifications do not contradict to the
observable data, as well as to find observable signatures of a new model.

Since 1950s~\cite{YM}, one uses the gauge symmetry as a main principle for
constructing interactions between elementary particles.  It is known, however,
that in theories with chiral couplings (i.e. when left and right-handed
fermions have different charges with respect to a gauge group), one may
encounter a situation when the classically conserved gauge current acquires an
anomalous divergence at one loop.  It is known that such anomalies destroy
consistency of the theory, making it
non-unitary~\cite{gross-jackiw}.\footnote{This represents a striking contrast
  with anomalies of \emph{global} symmetries, where global current
  non-conservation accounts for a new phenomena, not seen in the tree-level
  Lagrangian. The examples of this sort are the fast decay rate of $\pi^0$
  meson into two $\gamma$-quanta (famous ABJ-anomaly~\cite{abj}), the baryon
  and lepton number non-conservation in electroweak theory~\cite{thooft}, and
  proton decay in the presence of magnetic monopole~\cite{rubakov,callan}.}
In the Standard Model, where the couplings of fermions are chiral, charges of
all particles can be chosen to ensure the absence of all gauge
anomalies~\cite{sm-anom}. It may happen, however, that anomaly-free theory
looks like an anomalous one below certain energy scale.  An example of such a
theory would be an electroweak theory at the energy scale below the mass of
top-quark. Then anomaly is canceled by means of Wess-Zumino like
terms~\cite{df}.

Among many possible extensions of the Standard Model there are theories with
extra dimensions.  Their main idea is that our 4-dimensional physics is
embedded into a theory in a higher-dimensional space.  The Standard Model
fields are then realized as zero modes of a higher-dimensional ones. (An
example is given by an approach, often dubbed \emph{brane-world} models, where
zero modes of both matter and gauge fields are localized on the 4-dimensional
defect, called \emph{brane}~\cite{rub-shap,akama}).  Once our 4-dimensional
world is just a low-energy sector of a bigger theory, the anomaly analysis
changes drastically. For example, if a full theory is anomaly free (say, it is
formulated in an odd number of space-time dimensions, where all interactions
are vector-like), then one has no reason to expect separate anomaly
cancellation for the brane fields (for an explicit example of a brane-world
theory with an anomaly on the brane, see~\cite{Shap.qed}).  If the theory on
the brane is anomalous, then there is a specific type of \emph{brane-bulk
  interaction}. It is described by Chern-Simons-like terms in the low-energy
effective action in the bulk.  These terms are not gauge invariant in the
presence of a brane.  Therefore, they generate currents, flowing to the brane
and thus ensuring the gauge invariance of the full theory. Such mechanism is
known for a long time and is often called \emph{anomaly
  inflow}~\cite{fs,ch}.\footnote{It is interesting to note that already the
  authors of~\cite{ch} had mentioned in their paper that it would be
  interesting to apply their ideas to the brane-world setups
  like~\cite{rub-shap}.}
It is quite general and appears in many (physically very distinct) problems:
in quantum Hall effect~\cite{wen,frohlich},\footnote{A quantum Hall
  system~\cite{girvin} is a lower-dimensional example of the present problem.
  Its boundary can be considered as a $1+1$ dimensional ``brane'', embedded
  into $2+1$ dimensional bulk. Effective theory in the bulk is described by
  the $U(1)$ Chern-Simons theory, while on the boundary chiral excitations are
  localized. Conformal field theory describing these excitations was first
  derived from anomaly consideration~\cite{wen,frohlich}.}
in various field theories with solitonic objects in them (see
e.g.~\cite{naculich,hr,bhr}), in string and
M-theory~\cite{witten,I-brane,fhmm,bk}.  Thus, in the brane-world models a
number of questions arises:
{\parskip=0pt
  \begin{enumerate}
    \itemsep=0pt
  \item\label{item:1} Can one distinguish between a purely 4-dimensional
    anomaly-free theory, and a theory (also anomaly-free!) in which anomaly is
    canceled by a small inflow from extra dimensions? What new observable
    effects appear in the latter situation?
    
  \item \label{item:4} Is it possible to discriminate at low-energies between
    two different completions of anomalous 4-dimensional effective theory: the
    one, where anomaly is canceled by inflow from higher dimensions and purely
    4-dimensional one, where at higher energies there exist additional
    particles, which ensure anomaly cancellation (c.f.~\cite{df}).
    
  \item \label{item:2} For generic values of hypercharges the minimal Standard
    Model is anomalous.  However, for any hypercharges, electrodynamics is
    still vector-like.  Can this anomaly nevertheless be observed if this
    anomalous theory is embedded into a consistent higher-dimensional one?

  \item \label{item:3} What are experimental constraints for the parameters
    describing anomaly mismatch? Can the above observations serve as tests in
    the search for extra dimensions?
 \end{enumerate}
} %
\noindent In this paper we will attempt to answer the
questions~\ref{item:1}--\ref{item:2}. The question~\ref{item:3} will
be addressed in our next paper~\cite{anomaly-exp}. We would like to
note also that in this paper (as well as in the subsequent
paper~\cite{anomaly-exp}) we study only gauge anomalies.  The global
chiral anomalies in brane-worlds and corresponding complicated vacuum
structure of gauge fields has been discussed in~\cite{theta-vacuum}.

We consider a theory on a brane, which is anomalous, while the bulk effective
action possesses a Chern-Simons-like term, responsible for an inflow current.
A naive answer to the question~(\ref{item:1}) then would seem to be the
following: as anomaly means gauge current non-conservation, one could expect
that particles escape from the brane to extra dimensions. This would look like
a loss of unitarity from a purely 4-dimensional point of view, however
higher-dimensional unitarity would still be preserved. Anomalous current then
is simply a flow of such zero-mode particles into the bulk.  However, this
answer is incorrect.  Indeed, phenomenologically acceptable brane-world
scenario must have a mass gap between zero modes and those in the bulk for
both fermions and non-Abelian gauge fields~\cite{rubakov-rev,r-ds}.
Alternatively, the rate of escape of the matter from a brane must be
suppressed in some way (see e.g.~\cite{dubovsky}).
As a result, the low-energy processes on the brane simply cannot create an
excitation, propagating in the bulk.  If no matter can escape the brane (or
flow onto it) -- what is then an inflow current, which should be non-zero even
for weak fields causing anomaly in four-dimensional theory?  One seems to be
presented with a paradox.

We show that in reality the inflow current by its nature is a \emph{vacuum} or
\emph{non-dissipative} current.\footnote{A well-known example of such current
  is that of the Quantum Hall effect.}
What we mean by that is the following. Usually in electrodynamics one needs to
create particles from a vacuum to generate an electric current.  Such a
process is only possible if the available energy is larger than the mass of
the particles in the bulk.  Therefore, weak electric fields on the brane
cannot induce a current, carried by the bulk modes.  However, anomaly inflow
current is different. First, it is perpendicular to the field, does not
perform a work and thus is not suppressed by the mass of a particle in the
bulk. Second, such a current is not carried by any real particles excited from
the vacuum, being rather a collective effect, resulting from a rearrangement
of the Dirac sea.  This simple observation is in fact very important.
Essentially, it means that anomaly inflow is a very special type of brane-bulk
interaction.

Namely, we show that in the presence of parallel magnetic and electric fields
the latter changes as if photon had acquired mass.  As a result the spatial
distribution of an electric field changes, in particular it appears outside
the plates of the capacitor. Similarly, an electric field of an point-like
electric charge changes from Coulomb to Yukawa, when the charge is placed in a
magnetic field. 
The dynamics of the screening of electric charge happens to be quite peculiar
and can be shown to lead to the temporary appearance of \emph{dipole moment}
of elementary particles.

At first, we conduct our analysis in $U(1)$ gauge theory with the chiral
couplings to a matter. One example of such a theory would be an
electrodynamics, where left and right-handed particles have different electric
charges. The masses of such particles can then be generated only via the Higgs
mechanism with a charged Higgs field, which means that a photon in such a
theory becomes massive.  A different example is a theory, where axial current
comes from a neutrino with a small electric charge. In such theory photon
would remain massless.

Then we analyze the version of the Standard Model where electric charges of
electron and proton differ.\footnote{The electric neutrality of matter
  provides restrictions on this charge difference~\cite{charge-diff}. However,
  even under these restrictions the discussed effects can be pronounced
  enough to be observed~\cite{anomaly-exp}.} %
We show that although low-energy electrodynamics in this theory is
vector-like, there exist effects, similar to those, described above.

We would like to stress again that in the theories with extra dimensions there
is no immediate reason to expect that four-dimensional anomaly is so small as
it follows already from present experimental constraints on the corresponding
parameters. The effects which would follow from the existence of an anomaly
inflow are observable, but quite peculiar. They do not change drastically
physical content of the theory and can be ruled out or constrained only
experimentally.  Thus in those brane-world scenarios, in which
\emph{higher-dimensional} theory is automatically anomaly free, we are facing
yet another fine-tuning puzzle.

For distinctness we will often refer to a simple brane-world: the fermions in
the background of a kink, realizing a 4-dimensional brane, embedded into a
5-dimensional space as in~\cite{rub-shap} (although main effects are believed
to be model-independent).  We localize gauge fields in a spirit
of~\cite{oda,dubovsky,st,laine}, introducing an exponential warp-factor.  Such
warp-factor can arise dynamically because of the fermionic zero
modes~\cite{dvali-localization}.  Zero modes of both types of fields are
separated by mass gaps from the corresponding bulk modes. 

The paper is organized as follows. We begin
section~\ref{sec:qed} with describing a model of chiral electrodynamics,
embedded into a 5-dimensional $U(1)$ theory and analyze the microscopic
structure of the inflow and details of anomaly cancellation.  
Next, we turn in Section~\ref{sec:anom-sm} to the question of possible
extensions of the Standard Model, such that it would admit an anomaly.  We
consider one-parameter family of such modifications, with the free parameter
being the difference of (absolute values of) electric charges of proton and
electron.  In such model particles also acquire an anomalous dipole moment. We
discuss it in the Section~\ref{sec:proton}.  We conclude with a discussion of
future extensions of this work and some speculations.

Finally, we would like to stress that in this paper we conducted analysis on
the level of classical equations of motion. We discuss some aspects of quantum
theory and the validity of our approach in~\ref{sec:disc}.

\section{Anomalous electrodynamics and its observational\\ signatures.}
\label{sec:qed}

\subsection{5-dimensional electrodynamics in the background of a kink}
\label{sec:qed5d}

We start our analysis considering in this section the simplest model which
nevertheless catches main effects we are interested in.\footnote{In the next
  section we proceed with the similar consideration for the Standard Model.}
Namely, we will discuss a four-dimensional $U(1)$ theory with anomalous chiral
couplings to the fermions.  Such a theory suffers from a gauge anomaly,
meaning that on the quantum level its unitarity is
lost~\cite{abj,gross-jackiw,sm-anom,thooft}. Thus it cannot describe physics
of a 4-dimensional world. However, if one thinks about such an anomalous
theory as living on a brane, embedded in a 5-dimensional theory, situation
changes.  Five-dimensional theory is anomaly-free (all couplings there are
vector-like) and therefore there should exist a special type of interaction
(\emph{anomaly inflow}~\cite{fs,ch}) between the theories on the brane and in the
bulk, which ensures the unitarity and consistency of the total system.  Below
we demonstrate this mechanism in details.

For definiteness we will illustrate our steps by using a concrete model of
localization of both fermions and gauge fields.
Consider the following theory:\footnote{
  Our conventions are as follows: Latin indices $a,\dots,e=0,\dots,4$, Greek
  $\mu,\nu= 0,\dots,3$. We are using \emph{mostly negative} metric. Our brane
  is stretched in $0,\dots,3$ directions and is located at $x^4 = 0$. We will
  often use notations $t,x,z$ instead of $x^0,x^1,x^4$ and choose polar
  coordinates $(r,\theta)$ in
  the plane $(x^2,x^3)$.\label{fn:1}} %
\begin{equation}
  \label{eq:1}
  S = -\frac1{4\e^2}\int d^5 x\, \Delta(x^4) F_{ab}^2 + \int d^5
  x\,\sum_{f=1}^2 
  \bar\Psi_f(x)\Bigl(i{\Dsl}_f+m_f(x^4)\Bigr) \Psi_f(x) \:.
\end{equation}
Here $\e$ is a five-dimensional charge, with the dimensionality of
$(\mathrm{length})^{\frac12}$. There are two fermions $\Psi_{1,2}$,
interacting with the gauge fields with the different charges: ${\Dsl}_f = \psl
+ \frac{\e_f}\e \Asl,\,\e_1 \neq \e_2$.  The fermionic mass terms
$m_1(x^4)=-m_2(x^4)$ have a ``kink-like'' structure in the direction $x^4$,
realizing the brane: $m_1(x^4 \to \pm \infty)\to \pm m_\psi$. As it is
well-known (see e.g.~\cite{rub-shap}), in this model one has chiral fermion
zero modes localized on the brane and a continuous spectrum in the bulk,
separated from the zero mode by the mass gap $m_\psi$.  Localization of the
zero mode of gauge field is achieved by introduction of a warp-factor
$\Delta(x^4)$. Whenever $\int dx^4\Delta(x^4)$ is convergent the system has a
zero mode with the constant profile in the 5th direction.  Warp-factor
$\Delta(x^4)$ looks like an $x^4$ dependent coupling constant in the
action~(\ref{eq:1}), getting infinite as $|x^4|\to\infty$. The lattice study
of~\cite{laine} has confirmed that the perturbative and non-perturbative
analysis of (\ref{eq:1}) give the same result for the zero mode up to the
energies of the order of the mass gap, at least for $2+1\to 1+1$ reduction. We
will assume that this is the case for $4+1\to 3+1$.  Depending on the
warp-factor, the spectrum of gauge field may be discrete or continuous with or
without the mass gap between the zero and bulk modes~\cite{oda,dubovsky,st}.
We will often use $\Delta(x^4)=e^{-2\M \left|x^4\right|}$ in which case the
zero mode is separated from the bulk continuum by the mass gap $\M$~\cite{st}.
The nature of the warp-factor $\Delta(x^4)$ is not essential here, it can
originate from that of the metric or can arise dynamically, coming from
fermionic zero modes~\cite{dvali-localization}.

A low-energy effective action for the gauge fields of the theory~(\ref{eq:1})
will contain terms, describing interaction between the brane and the bulk:
\begin{equation}
  \label{eq:2}
  S_{\text{5d}} = -\frac 1{4\e^2}\int d^5 x\, \Delta(x^4)\, F^2_{ab}  + S_{\cs} +
  S_{\zm}\, .
\end{equation}
The origin of two last terms is the following.  The computation of the
fermionic path integral of our theory is separated into the 
determinant of the 5-dimensional bulk modes of the fermions and
that of the 4-dimensional chiral fermionic zero modes.  In the former case one
can show (essentially repeating the computations of~\cite{redlich}), that
after integrating out the non-zero modes of the fermions, the 5-dimensional
effective action of the model~(\ref{eq:1}) acquires an additional
contribution of Chern-Simons type
\begin{equation}
  \label{eq:3}
  S_\cs = \frac14\int dx^5\,\kappa(x^4)\epsilon^{abcde}A_a F_{bc} F_{de}\,. 
\end{equation}
Compared to the usual case~\cite{redlich,a-gd-pm}, Chern-Simons term in models
like~(\ref{eq:1}) acquires additional factor $\kappa(x^4)$.  It is proportional to the difference of the charges of the
five-dimensional fermions and depends on the profile of the domain wall. In
the limit of an infinitely thin wall and the large fermionic mass gap
$\kappa(x^4) = \kappa_0\sign(x^4)$.  Below we sketch the derivation of this
function.

Terms like~\eqref{eq:3} generically appear in the theories with anomaly
inflow~\cite{ch,naculich,I-brane,fhmm,bk,hr,bhr}. In situations with more than
one extra dimension an effective action will have similar structure, with the
fifth coordinate being suitably chosen radial coordinate.  Such terms can also
appear for non-Abelian fields.

Evaluating of the determinant of the 4-dimensional Dirac operator in the
background of the gauge field (and, possibly, Higgs field, see below), we get
$S_\zm$.
Let us denote the \emph{4-dimensional} charges of the left and right-handed
fermionic zero modes as $e_L/e$ and $e_R/e$.  They are proportional to their
5-dimensional counterparts $\e_1$ and $\e_2$ and thus $e_L \neq e_R$.  As a
result $S_\zm$ is \emph{not gauge invariant}. The manifestation of this fact
is the anomalous divergence of the gauge current at one-loop level:
\begin{equation}
  \label{eq:5}
  \p\mu \la j^\mu_\zm\ra = \frac{e_R^3 - e_L^3}{16\pi^2 e^3}
  \epsilon^{\mu\nu\lambda\rho}F_{\mu\nu}F_{\lambda\rho}\,\delta(x^4)\,,
  \qquad\text{where}\quad \la j^\mu_\zm \ra\equiv \frac{\delta S_\zm}{\delta
    A_\mu} \: .
\end{equation}
The action~(\ref{eq:3}) is also non-gauge invariant, and the variations of
these two terms cancel each other (as we will demonstrate below). Divergence
of the current~(\ref{eq:5}) is proportional to the transversal profile of the
fermion zero modes. In this paper we will be interested in the characteristic
energy scales much less than the mass gap of the gauge fields $\M$, which is
in turn much less than the fermionic mass gap: $\M\ll m_\psi$. Therefore the
profile of the fermion zero mode can be approximated by the delta-function
$\delta(x^4)\approx 2m_\psi e^{-m_\psi|x^4|}$, which appears in
eq.~(\ref{eq:5}).

Equations of motion coming from the action~(\ref{eq:2}) are:
\begin{equation}
  \label{eq:6}
  \frac1{\e^2}\p b \Bigl(\Delta(x^4)F^{\mu b}\Bigr) = J^\mu_\cs + \la
  j_\zm^\mu\ra,\qquad a,b=0,\dots,4
\end{equation}
and
\begin{equation}
  \label{eq:7}
 \frac{1}{\e^2}\Delta(x^4)\p\mu F^{4 \mu}  = J^4_\cs, \qquad \mu=0,\dots,3.
\end{equation}
In the right hand side of eq.~(\ref{eq:6}) there is a current of the zero
modes $j_\zm^\mu$ (we will omit brackets $\la\dots\ra$ from now on).  The
variation of the effective action $S_\zm$ with respect to the $A_4$ is equal
to zero.\footnote{Indeed, $j^4_\zm \sim \la\e_1\bar\Psi_1 \gamma^5\Psi_1 +
  \e_2\bar\Psi_2\gamma^5\Psi_2\ra$. Zero modes of $\Psi_1$ and $\Psi_2$ have
  definite 4-dimensional chirality.  Therefore $\bar\Psi_1\gamma^5\Psi_1 =
  -\bar\Psi_L \Psi_L = 0$ and analogously for $\Psi_2$.}

The current $J_\cs$ in the right hand side of eqs.~(\ref{eq:6})--(\ref{eq:7})
is defined via
\begin{equation}
  \label{eq:8}
  J^a_\cs(x^a) = \frac{\delta S_\cs}{\delta A_a} = 
  \frac{3}{4}\kappa(x^4)
  \epsilon^{abcde}F_{bc}(x^a)F_{de}(x^a)+\frac12\kappa'(x^4)\delta^{a}_\mu 
\epsilon^{\mu\nu\lambda\rho} A_\nu F_{\lambda\rho},\quad
  \kappa'(x^4) =\frac{d \kappa(x^4)}{d x^4}\:.
\end{equation}
Note that in the right hand side of eq.~(\ref{eq:8}) there is a term,
proportional to the derivative of $\kappa(x^4)$. This term is non-zero only in
the vicinity of the brane (as $\kappa'(x^4)$ is proportional to the
delta-function of $x^4$). Thus, essentially it corresponds to the modification
of definition of the brane current $j^\mu_\zm$~(\ref{eq:5}).  This additional
term is responsible for making anomaly of the zero modes \emph{gauge
  invariant}~\cite{naculich,hr}, as the inflow current~(\ref{eq:5}) is always
gauge invariant in the $U(1)$ case.  Thus, terms proportional to
$\kappa'(x^4)$ correspond to the local counterterms, shifting between the
\emph{covariant} and \emph{consistent} anomalies~\cite{cov-anom}. In what
follows we will always consider covariant anomaly of the brane modes and
ignore terms in the bulk, proportional to the $\kappa'(x^4)$. The Chern-Simons
current $J^a_\cs$ is a non-dissipative one as it does not perform a work due
to its antisymmetric structure.  In addition to that it does not depend on the
mass of the fermions in the bulk and hence is not suppressed by the mass gap.
If a gauge field has four-dimensional components with
$\epsilon^{\mu\nu\lambda\rho}F_{\mu\nu}F_{\lambda\rho}\neq0$, there exists a
current $J^4_\cs$, flowing onto the brane from the extra dimension~\cite{ch}.

Trivial consequence of eqs.~(\ref{eq:6})--(\ref{eq:7}) is
\begin{equation}
  \label{eq:9}
  \p a J_\cs^a + \p\mu j_\zm^\mu = \frac1{\e^2}\p a\p b F^{ab} = 0\:.
\end{equation}
It means that if coefficients in front of the $S_\cs$ and $S_\zm$ were not
correlated, by virtue of~(\ref{eq:5}) the solutions of eq.~\eqref{eq:9} would
be field configurations with
$\epsilon^{\mu\nu\lambda\rho}F_{\mu\nu}F_{\lambda\rho}=0$. However, to
preserve the gauge-invariance in the action~\eqref{eq:2} the divergences of
currents in the left hand side of~\eqref{eq:9} should be equal.  From the
definition~(\ref{eq:8}) one can see that
\begin{equation}
  \label{eq:10}
  \p a J^a = \frac34\kappa'(x^4)
  \epsilon^{\mu\nu\lambda\rho}F_{\mu\nu}F_{\lambda\rho} = \frac32\kappa_0
  \epsilon^{\mu\nu\lambda\rho}F_{\mu\nu}F_{\lambda\rho}\delta(x^4)\:.
\end{equation}
Comparing~(\ref{eq:10}) with eq.~(\ref{eq:5}) we can choose the coefficient of
$\kappa_0$ of the function $\kappa(x^4)$ so that the current $J_\cs$ cancels
the anomalous divergence of the current $j_\zm^\mu$ and make the
theory~(\ref{eq:2}) consistent and gauge-invariant. Namely
\begin{equation}
  \label{eq:11}
  \kappa(x^4) = \kappa_0\sign(x^4),\quad\text{where}\quad \kappa_0 \equiv
  \frac{e_L^3 - e_R^3}{24\pi^2e^3} \:.
\end{equation}
Microscopical calculations of $\kappa(x^4)$ confirm that this is indeed the
case.

The relation between 5-dimensional fields $F_{\mu\nu}$, entering
eqs.~(\ref{eq:6})--(\ref{eq:7}) and the 4-dimensional ones is given by:
\begin{equation}
  \label{eq:12}
  \frac 1{e^2} \bF_{\mu\nu} =\frac 1{\e^2}\int^{\infty}_{-\infty} dx^4
  \Delta(x^4) F_{\mu\nu}\:. 
\end{equation}
Here $e$ is a \emph{four-dimensional $U(1)$ coupling
  constant}, related to the 5-dimensional one via
\begin{equation}
  \label{eq:13}
  \frac 1{e^2} =\frac 1{\e^2}\,\int dx^4 \Delta(x^4)
\end{equation}
(to see that, one should substitute $F_{\mu\nu}$, independent on the 5th
coordinate, into the kinetic term in~(\ref{eq:2}) and integrate over $x^4$).
Field $\bF_{\mu\nu}$ satisfies the 4-dimensional Maxwell equations $\p\nu
\bF^{\mu\nu}=e^2\bj^\mu$ with the current in the right hand side given by:
\begin{equation}
  \label{eq:14}
  \bj^\mu=\int^{\infty}_{-\infty} dx^4 \,\left(j^\mu_\zm + J^\mu_\cs -
    \p4\Bigl(\Delta(x^4) F^{\mu4}\Bigr)\right)\:.
\end{equation}
This current is conserved: $\p\mu \,\bj^\mu=0$ as a consequence of
eqs.~(\ref{eq:7}) and~(\ref{eq:9}).

\begin{figure}[t]
  \centering \includegraphics[scale=.5]{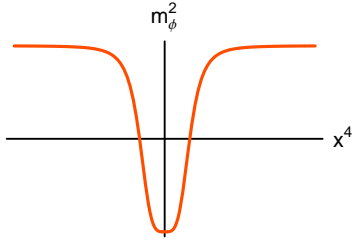}
  \caption{Profile of the mass of the Higgs field.}
  \label{fig:higgs}
\end{figure}
In the theory~(\ref{eq:1}) fermionic zero modes on the brane are massless. To
make this model more realistic, we should add mass to these zero modes as
well.  The only way to make the electro-dynamics anomalous is to take left and
right moving fermions with different electric charges. Thus one can only
introduce a mass term via the Higgs mechanism with an electrically charged
Higgs field, i.e. to add the following term to the bulk
Lagrangian~(\ref{eq:1}):
\begin{equation}
  \label{eq:16}
  S_\phi = \int d^5x\left[ \bigl|D_a \phi\bigr|^2 -
    m_\phi^2 (x^4)|\phi|^2  -\frac{\lambda}4   |\phi|^4+
    f\bar \Psi_1\Psi_2 \phi +
  \mathrm{h.c.} \right]\:,
\end{equation}
where $D_\mu \phi = i\p\mu \phi + (\frac{e_L}e - \frac{e_R}e) A_\mu \phi$ and
the Higgs mass $m_\phi^2(x^4)$ is negative at $x^4=0$ and tends to the
positive constant in the bulk, as $|x^4|\to \infty$ (Figure~\ref{fig:higgs}).
Thus Higgs field has a non-trivial profile $\phi(x^4)$, with $\phi(0)\neq 0$
and $\phi(x^4)\to 0$ at $x^4\to\pm \infty$.  So the symmetry is restored far
from the domain wall. The Higgs mass far from the brane is taken to be large,
so that the scalar is localized on a brane.  The Higgs field has an
expectation value, which on the brane is $\la\phi(0)\ra\sim v$ (we take the
scalar's self-coupling $\lambda$ to be large, so that Higgs mass $m_H$ is much
bigger than $fv$ -- Yukawa mass of the fermions on the brane).  As the Higgs
field is charged, the photon should acquire mass for $x^4\approx 0$). To
determine it exactly one needs to find the spectrum of the theory with action
given by the sum of~(\ref{eq:1}) and~(\ref{eq:16}), but for us it is only
important that $m^2_\gamma \sim \kappa_0^2$, where parameter $\kappa_0$ was
defined in~(\ref{eq:11}).  In Section~\ref{sec:anom-sm} we will analyze the
consequences of anomaly inflow in the Standard Model, where Yukawa mass of the
fermions does not necessarily lead to the massive photon.

Dynamics of the Higgs field $\phi$ simplifies in the limit of the large
(compared to the energy scale of the processes on the brane) VEV $v$.  In this
limit it
is convenient to work in the \emph{unitary gauge}, in which the phase of the
Higgs field is fixed (e.g. via $\phi=|\phi|$). In the regions of space where
the gauge field has non-trivial $F_{\mu\nu}$, it is impossible to have
$|\phi|=\const$~\cite{amb-Higgs}. However, variations of the $|\phi(x^a)|-v$
are suppressed by the Higgs's VEV, we can neglect them and take
$\phi=v=\const$ on the brane. Now the vector field is massive and longitudinal
component becomes a physical degree of freedom. From equations of motion one
can obtain the following condition on the field $A_a(x^a)$:
\begin{equation}
  \label{eq:17}
    \p a \bigl(m^2_\gamma(x^4) A^a\bigr) + \p\mu j^\mu_\zm + \p a
    J^a_\cs  =0\:.
\end{equation}
Eq.~(\ref{eq:17}) is just an expression of the gauge invariance of the total
action. Divergences of $j_\zm$ and $J_\cs$ precisely cancel each
other~(\ref{eq:9}).  Therefore eq.~(\ref{eq:17}) implies that
\begin{equation}
  \label{eq:18}
  \p a \bigl(m^2_\gamma(x^4) A^a\bigr)  =0 \:.
\end{equation}

Once we have added the mass to the zero mode fermions, we can integrate them
out and obtain effective action $S_\zm$~(\ref{eq:2}), written in unitary
gauge. The variation of this action with respect to $A_\mu$ gives gauge
invariant current~\cite{df}:
\begin{equation}
  \label{eq:19}
   j^\mu_{\zm} = 3\kappa_0 \epsilon^{\mu\nu\lambda\rho}\frac{\phi^*
     \overleftrightarrow D_\nu\phi}{v^2} F_{\lambda\rho}\delta(x^4) \:.
\end{equation}
We will often call this current \emph{the D'Hoker-Farhi current} denote it
$j^\mu_\df$. In the unitary gauge eq.~(\ref{eq:19}) becomes
\begin{equation}
  \label{eq:20}
  j^\mu_{\df} = 3\kappa_0 \epsilon^{\mu\nu\lambda\rho}A_\nu
  F_{\lambda\rho}\delta(x^4)  \:.
\end{equation}
This current has a divergence, equal to that of the Chern-Simons current
($\p\mu j^\mu_\df = - \p a J^a_\cs$):
\begin{equation}
  \label{eq:21}
  \p\mu j^\mu_\df = \frac32\kappa_0
  \epsilon^{\mu\nu\lambda\rho}F_{\mu\nu}(x^\mu,x^4)F_{\lambda\rho}(x^\mu,x^4)
  \delta(x^4)     \:.
\end{equation}

\par
A purpose of this work is to explore in details this anomaly cancellation
mechanism.  Namely, \emph{what constitutes the inflow}? Zero modes of fermions
are confined to the brane and cannot propagate in the bulk, while bulk modes
are massive and cannot be excited at low energies. Another question is
\emph{whether 4-dimensional observer can distinguish} between an anomaly-free,
intrinsically 4-dimensional theory and 5-dimensional (anomaly-free!)
brane-world theory, in which a very small anomaly of the 4-dimensional
fermions on the brane is canceled by an inflow from the bulk.

\subsection{Anomalous electric field of a capacitor}
\label{sec:observ-inflow}

As we have seen, on the microscopic level anomaly
inflow is a collective effect of reorganization of
the Dirac see of the massive fermions in the bulk in the presence 
of  a field configuration with non-zero
$\epsilon^{\mu\nu\lambda\rho}F_{\mu\nu}(x^a)F_{\lambda\rho}(x^a)|_{x^4=0}\sim
(\vec E\cdot \vec H)$.

The simplest way to obtain non-zero $\vec E\cdot \vec H$ is to choose both $E$
and $H$ to be parallel to each other and constant (in $x^\mu$) in some region
on the brane.  We realize such fields by two parallel infinite plates in the
vacuum, separated in the direction $x^1$ by the distance $2d$ -- a capacitor
with initial charge densities $\pm\sigma_0$ on plates, placed in the magnetic
field. Magnetic field is created by an (infinite in the $x$-direction)
solenoid with the radius $R$.

Before these fields are turned on, the vacuum structure on the classical level
is unchanged and there is no difference with non-anomalous theory.  Also, the
effects of anomaly are controlled by the parameter $\kappa_0$, which is
naturally assumed to be very small. That is why the most natural setup to
study the anomaly inflow is qualitatively the following.  We assume that the
vacuum current is zero and an electric field is equal to its value for
non-anomalous theory. Then we turn on a parallel magnetic field and expect
that small (perturbative in $\kappa_0$) inflow currents start to change the
charge distribution on the brane trying to compensate the anomalous field
configuration. According to this picture, in Section~\ref{sec:perturb} we will
choose initial condition describing the absence of anomalous vacuum structure
and will consider an initial stage of inflow (linear in time) perturbatively,
in the first order in $\kappa_0$.

We will see, however, that this naive approach does not work and perturbation
theory breaks down. Therefore, we will have to study the full non-linear
equations non-perturbatively in $\kappa_0$. Then, as we will show in
Section~\ref{sec:final-state}, a non-analytic in $\kappa_0$ static solution
exists. The properties of this solution will define the main observable effect
of the anomaly inflow, studied in the present paper. It is, nevertheless,
instructive to study the dynamical problem to understand qualitatively how
this state is formed.  As we will see in Section~\ref{sec:old-part}, from the
4-dimensional point of view the picture of the inflow given by perturbation
theory is qualitatively correct and knowing the non-perturbative static
solution one can define it properly.

\subsubsection{Linear stage of anomaly inflow: naive perturbative treatment}
\label{sec:perturb}

As in general we expect the parameter $\kappa_0$ to be extremely small, we
will try to solve the Maxwell equations~(\ref{eq:6})--(\ref{eq:7}) by
perturbation theory in $\kappa_0$. To find linear in time corrections to the
fields at $t=0$, we should specify the initial values of all fields. Our
initial conditions should satisfy the Gauss constraint\footnote{In this
  Section we use notations $t\equiv x^0$, $x\equiv x^1$, $z\equiv x^4$ and
  choose polar coordinates $(r,\theta)$ in the plane of the capacitor
  $(x^2,x^3)$ (c.f.  footnote~\vref{fn:1}).}
\begin{equation}
  \label{eq:22}
  \frac1{\e^2} \Bigl(\p z\bigl(\Delta(z)E^z\bigr) +\Delta(z) \div \vec E\Bigr)
  = J^0_\cs + j_\df^0 + \sigma_0\delta(z) \Bigl(\delta(x+d) -\delta(x-d)\Bigr)\:,
\end{equation}
where $J^0_\cs$ and $j^0_\df$ are expressed in terms of components $F_{ab}$
and are both proportional to the $\kappa_0$ (c.f.~(\ref{eq:8})
and~(\ref{eq:20}) correspondingly).
Because we have added a charged Higgs field~(\ref{eq:16}) and the
electrodynamics became massive, specifying initial values of $F_{ab}$ is not
enough and one should also set values of the gauge potential at $t=0$. We
choose all but $A_\theta$ components $A_a$ initially equal to zero (see
Appendix~\ref{app:linear} for details). The time derivatives of $A_a$ can then
be extracted knowing electric fields at $t=0$ and using
condition~(\ref{eq:18}).  In this paper we restrict our analysis only to such
initial conditions.  In general initial conditions on the vector field $A_a$
will depend on how one turns on electric and magnetic fields.  Corresponding
analysis (similar to that of~\cite{amb-anomaly}) will be presented elsewhere.

In addition to these initial condition we will choose as a zeroth
approximation such configuration of electromagnetic fields $F_{ab}$ that
$J^0_\cs\bigr|_{t=0}=0$. The reason for that is as follows: from~(\ref{eq:8})
one has $J^0_\cs =\frac34 \kappa(z) F_{xz} H^x$. In the absence of anomaly
$F_{xz}$ would be equal to zero for our configuration of charges. Therefore,
it is natural to assume that initially $F_{xz}$ is equal to zero in the full
theory as well.  Initial conditions $J^0_\cs = 0$ and $j^0_\df=0$ mean that we
should prepare an electric field configuration, satisfying 5-dimensional Gauss
constraint in the theory with $\kappa_0=0$ (we will mark these fields by the
symbol $^\0$).  Then we find (perturbative) corrections to these fields due to
anomalous current and analyze effects arising on the initial (linear) stage of
inflow.  The details of the computations can be found in
Appendix~\ref{app:naive}, here we only quote the main results.

With our choices for the fields $F_{ab}^\0$, initially only time component of
the D'Hoker-Farhi current changes with time (see Appendix~\ref{app:naive}):
\begin{equation}
  \label{eq:35}
  \p t j^0_\df(x^\mu)\biggr|_{t=0}\hskip -1ex  = -6\kappa_0 E_x^\0(x,z)
  H_x^\0(r,z) 
  \delta(z) \:,
\end{equation}
If the theory were purely four-dimensional, eq.~(\ref{eq:35}) would have
described the rate of anomalous particle production, essentially the measure
of non-unitarity of the theory.  In the 5-dimensional brane-world the
interpretation of eq.~(\ref{eq:35}) is quite different.  Namely, one should
not think about the inflow current, as ``bringing particles to the brane''.
This is obvious as all the light (zero mass) particles are confined to the
brane and do not propagate in the bulk, while the fermions, which live in the
bulk are too massive to be excited at low energies.  Thus, inflow
current~(\ref{eq:8}) is essentially a \emph{vacuum current} -- redistribution
of the particle density without actual creation of the charge carriers. This
fact can be checked by the direct microscopic computations of the vacuum
average of $\bra 0 J_\cs^z\ket 0$ in the full theory~(\ref{eq:1}).  The
divergence $\p a J_\cs^a$ is non-zero only in the position of the brane $z =
0$ and thus modifies the charge density only there (which is reflected by the
delta-function in eq.~\eqref{eq:35}). One may think of the effect of the
vacuum current as of a dielectric susceptibility of the vacuum of
four-dimensional theory, embedded in the 5-dimensional space-time.

Notice, that since there is no $J^0_\cs$, eq.~(\ref{eq:35}) means that
\emph{the total electric charge density changes with time}.  If the charge
density changes in the finite volume on the brane, this leads to the change of
both $E^x$ and $E^z$. (If initial field $E^x$ and $H^x$ uniformly filled all
the brane and were localized in $z$ direction, then the appearance of the
anomalous density would not lead to the change of the $x$ component of the
electric field. The change in $E^z$ would be unobservable, since this field is
antisymmetric in $z$ and so $\int dz\, \Delta(z)E^z = 0$).  For the case when
$\vec E\cdot \vec H$ is non-zero only in the finite volume on the brane we
explicitly find the change in $E^x$ by solving Maxwell
equations~(\ref{eq:6})--(\ref{eq:7}):
\begin{equation}
  \label{eq:109}
    E_x^\1(x^a) = -6\e^2\frac{\kappa(z)}{\Delta(z)}
    E_z^\0(x,z)H^\0_x(r,z)t+\CO(t^2) \:.
\end{equation}
Now one can see that \emph{our perturbation theory approach has a problem}.
Indeed, although $E_x^\1$ is proportional to $\kappa_0$, it grows in $z$ as
$\frac1{\Delta(z)}$ while $E_x^\0$ decays as $|z|\to \infty$. The correction
term~(\ref{eq:109}) becomes larger than $E_x^\0$ at $|z|\sim -\frac{\log
  \kappa_0}{\M}$. Eq.~(\ref{eq:109}) gives correction to the
\emph{5-dimensional} electric field. The corresponding 4-dimensional field is
obtained via~(\ref{eq:12}). As shown in Appendix~\ref{app:naive}, the integral
in~(\ref{eq:12}) is saturated at $|z|\sim Mx^2$ where characteristic $|x|\gg
\frac1{\M}$. Therefore in the physically relevant region
\begin{equation}
  \label{eq:44}
  E_x^\1 > E_x^\0\quad \text{for}\quad |z|\gtrsim -\frac{\log \kappa_0}{\M} \:.
\end{equation}
Thus, we see that the naive form of perturbation theory does not work. It
means that something was wrong with our qualitative picture of the inflow and
that we have to study full equations non-perturbatively.

\subsubsection{Analysis of a static solution}
\label{sec:final-state}

We have seen in the previous section that effects, caused by anomaly inflow
are non-perturbative. As the full system of non-linear Maxwell equations is
too complicated to solve exactly, we will compute a static solution.  This
means that we do not study how anomaly inflow modifies the initial fields, but
rather describe a final state of the inflow. We will consider the simplest
case: a capacitor with infinite plates, when these equations are reduced to
the $2+1$ ones.  Below we will only sketch the computations, for details
reader should refer to Appendix~\ref{sec:final-state-compute}.  We will return
to the perturbation theory in the Section~\ref{sec:old-part}.

For static solution we can describe the electric field in terms of the
electro-static potential $\Phi$: $E_i = -\p i \Phi$, $E_z = -\p z \Phi$.  Then
the Gauss constraint can be re-written as (see
Appendix~\ref{sec:final-state-compute} for details):
\begin{equation}
  \label{eq:45}
  \p z \Bigl(\Delta(z) \p z \Phi\Bigr) + \Delta(z) \nabla^2 \Phi = \frac{36\e^4
    \kappa_0^2}{r^2 \Delta(z)}\Bigl( F_{r\theta}^2 + F_{x\theta}^2 + F_{\theta
      z}^2\Bigr) \Phi+ \e^2 q(x)\delta(z)+j^0_\df\:,
\end{equation}
where $\nabla^2$ is a 3-dimensional Laplacian in the coordinates $x,r,\theta$.
Components $F_{r\theta},\, F_{x\theta}\, F_{z\theta}$ can all be expressed via
single function $A_\theta$ (because we choose all the gauge connection to be
$\theta$ independent).  The Maxwell equation~(\ref{eq:6}) for $\mu=\theta$
becomes then an equation for $A_\theta$:
\begin{equation}
  \label{eq:46}
  \p z \Bigl(\Delta(z) \p z A_\theta\Bigr) + \Delta(z) \nabla^2 A_\theta =
  \frac{36\e^4 \kappa_0^2 \Phi}{r^2 \Delta(z)}\Bigl(\p x A_\theta \,\p x \Phi  +
  \p z A_\theta\p z \Phi + \p r A_\theta \p r \Phi\Bigr)\:.
\end{equation}
We will analyze this solution as follows. First, we notice that if the right
hand side of eq.~(\ref{eq:46}) were identically zero, then there would be a
solution $F_{r\theta} = r H^x$, $H^x=\const\equiv \bH_0$, $F_{x\theta} =
F_{\theta z}=0$.  Then one can easily find an explicit solution of
eq.~(\ref{eq:45}) (under these assumptions Gauss law becomes identical to that
of the $2+1$ dimensional case with the point-particle as the source).  After
that we show that corrections to the found solutions of
eqs.~(\ref{eq:45})--(\ref{eq:46}) due to the non-zero right hand side
of~(\ref{eq:46}) are of the order $\kappa_0^2$. Note, that potentially the
first terms in the right hand sides of both~(\ref{eq:45}) and~(\ref{eq:46})
can be non-perturbative, as they are proportional to the $\frac1{\Delta(z)}$.
This is precisely why the naive perturbation theory of the
Section~\ref{sec:perturb} did not work.

After the Fourier transform in $x,r,\theta$-directions: ($\nabla\to i \vec p$)
and a substitution $\Phi=\frac{\psi_p(z)}{\sqrt{\Delta(z)}}$
equation~(\ref{eq:45}) becomes\footnote{We have neglected a D'Hoker-Farhi
  charge density $j^0_\df$ in eq.~(\ref{eq:47}) as compared to~(\ref{eq:45}).
  $j^0_\df$ is proportional to $\kappa_0 \delta(z)$, and therefore it can be
  treated perturbatively, as any function which is proportional to $\kappa_0$
  and does not have a growing profile in $z$ direction.}
\begin{equation}
\label{eq:47}
  (\hat H + p^2) \psi_p(z) = -\e^2 q_0 \delta(z)\:,
\end{equation}
where operator $\hat H$ does not depend on $p$:
\begin{equation}
\label{eq:48}
  \hat H = -\p z^2 + \Bigl(36\e^4\kappa_0^2 \bH_0^2\, e^{4\M|z|} + \M^2 -2
  \M\delta(z)\Bigr) \:.
\end{equation}
Let us denote the eigen-function of~(\ref{eq:48}) by $\psi_n(z)$ and its
eigen-values by $m_n^2$: $\hat H\psi_n(z) = m_n^2\psi_n(z)$.  One can show
(see Appendix~\ref{sec:final-state-compute}) that
\begin{equation}
  \label{eq:49}
  m_0^2 = 12 \kappa_0 \e^2 \bH_0 \M + \CO(\kappa_0^2\log\kappa_0) \approx
  12\kappa_0 e^2 \bH_0 \quad \text{and}\quad m_n^2 > \M^2\quad\text{for}\quad n>0
\end{equation}
(recall that $e^2$ is a \emph{four-dimensional coupling
  constant}~(\ref{eq:13})). Notice that $m_0$ depends only on the
four-dimensional quantities (electro-magnetic coupling constant $e^2$ and the
magnetic field $\bH_0$ measured by a 4-dimensional observer)! The
eigen-function for the eigen-value $m_0^2$ is given by
\begin{equation}
  \label{eq:50}
  \psi_0(z) = \frac{m_0}{\sqrt{2\pi\M}}K_{\nu}\left(\tfrac{m_0^2}{4\M^2}\, e^{2 M   |z|}\right)\quad \text{where} \quad \nu =
  \frac12\sqrt{1-\frac{m_0^2}{\M^2}}.
\end{equation}
Here $K_\nu(u)$ is a modified Bessel function of the second kind\footnote{
  For definition see e.g.~\textbf{8.40} in~\cite{gradshteyn}. In
  \texttt{Mathematica} this function is defined as \texttt{BesselK[n,x]}.}
and $c_0$ is a normalization constant.  The solution of eq.~(\ref{eq:45}) can
be easily found for any charge distribution $q(x)$.  For the point-particle in
the uniform magnetic field an electrostatic potential, created by the particle
will be of the Yukawa form:
\begin{equation}
  \label{eq:58}
  \Phi(x,r,z) = \frac{q_0}{4\pi}\frac{ e^{-m_0\sqrt{x^2+r^2}}}{\sqrt{x^2+r^2} }\chi_0(z)\:,
\end{equation}
where the mass $m_0$ depends on the anomaly coefficient and the magnetic field
$H^x$~(\ref{eq:49}). The profile of the solution~(\ref{eq:58}) in the $z$
direction $\chi_0(z)$ is proportional to $\psi_0(z)$ (see also
Appendix~\ref{sec:final-state-compute}):
\begin{equation}
  \label{eq:52}
  \chi_0(z) \equiv \frac{\psi_0(z)\psi_0(0)}{\sqrt{\Delta(z)}}
  \approx\frac{m_0}{\sqrt{6\pi\Delta(z)}} K_{\nu}(
  \tfrac{m_0^2}{4\M^2} e^{2\M|z|}) \:,
\end{equation}
(where in the last equality we assumed that $m_0 \ll \M$).  Function
$\chi_0(z)$ is sharply localized in the region
\begin{equation}
  \label{eq:53}
  |z| \lesssim 
  \frac1{\M}\log\frac {2\M}{m_0} =\frac1{2\M}\log\frac{\M^2}{3\, e^2 \kappa_0
    \bH_0} \:.
\end{equation}
Indeed, asymptotics of the Bessel function $K_\nu(u)$ is
$\frac{e^{-u}}{\sqrt{2\pi u}}$ for $u\gg 1$, therefore outside the specified
region potential $\Phi$ decays as an exponent $e^{-\Delta(z)}$. Thus, the
function $\Phi(x,z)$ exponentially decays in $z$ direction on the scale
proportional to the $\M^{-1}$ and depending on $\kappa_0$ (potentially very
small) only logarithmically, while the scale $m_0^{-1}$ of the exponential
decay of the potential in the $x$ direction does not depend on $\M$ and is
proportional to the $\sqrt{\kappa_0}$. As a result $m_0\ll \M$ for any
physically plausible values of magnetic field $\bH_0$.  Notice, that the
function $K_\nu(u)$ is non-regular for small $u$ and the same is true for
$\chi_0(z)$ as a function of $\kappa_0$.

Now it is easy to show that equation~(\ref{eq:46}) for $A_\theta$ only gives
small corrections to the constant magnetic field $\bH_0$. Indeed, the right
hand side of eq.~(\ref{eq:46}) is proportional to the $\kappa_0^2$. In case of
eq.~(\ref{eq:45}) such term was non-perturbative, because it was divided by
$\Delta(z)$ and thus could get arbitrarily large as $|z|\to \infty$. In the
case at hand, however, the right hand side is proportional to the
$\frac{\kappa_0^2 \Phi(x,r,z)}{\Delta(z)}$. As $\Phi$ decays in $z$ much
faster than $\Delta(z)$, this term is always small and therefore all the
corrections to the $A_\theta(r) = \frac 12 r^2 H^x$ are of the order
$\kappa_0^2$. Similarly, there are corrections to the $\Phi(x,r,z)$, which are
of the order $\CO(\kappa_0)$ due to the presence of D'Hoker-Farhi term
$j^0_\df$.

To summarize, we see from eq.~(\ref{eq:58}) that anomaly inflow causes
screening of an electric charge with the screening radius being $m_0^{-1}$.
This means, that total amount of anomalous charge, which inflows on the brane
is equal to an initial electric charge $q_0$ of a particle.

Similarly to eq.~(\ref{eq:58}), in case of capacitor with infinite plates one
can see that for $|x\pm d|\gg \M^{-1}$ the expression $\Phi$ is given by an
electro-static potential created by two infinite charged plates in 3 spatial
dimensions for a \emph{massive} electric field with the mass $m_0$:
\begin{equation}
\label{eq:51}
  \Phi(x,z) = \phi_0(x) \chi_0(z)=-\frac{\e^2 {\sigma_0}}{2m_0} \left(
    e^{-m_0 |x-d|} -    e^{-m_0 |x+d|}\right) \chi_0(z)
\end{equation}
(i.e. $\p x^2 \phi_0(x) - m_0^2 \phi_0(x)= q(x)$).  Constructing 4-dimensional
electric field from~(\ref{eq:51}) we get:
\begin{equation}
  \label{eq:54}
  \boxed{
    \bE^x(x) = -\frac{\bE_0}2  \left[\sign (x-d)e^{-m_0 |x-d|} -
      \sign(x+d) e^{-m_0 |x+d|}\right]
}\:. %
\end{equation}
Here $\bE_0$ is a value of 4-dimensional field, which would be created between
the plates of the capacitor in the theory without anomaly $\bE_0 = \sigma_0
e^2$. In the presence of anomaly distribution of an electric field changes.
Inside a capacitor it diminishes:
\begin{equation}
  \label{eq:55}
  \bE^x_{inside} = \bE_0 e^{-m_0 d}\cosh m_0 x,\quad |x| < d\:,
\end{equation}
but appears outside:
\begin{equation}
  \label{eq:56}
  \bE^x_{outside} = \bE_0 e^{-m_0 (|x| -d)}(1 - e^{-2 m d}), \quad |x|> d\:.
\end{equation}
As one can see, inside the capacitor the field is smaller, than would be in
the absence of anomaly, however, it appears outside the capacitor. If
$\kappa_0$ is such that $m_0 d\ll 1$, the field outside the capacitor (for $
\frac1{\M}\ll(x-d)\ll \frac1{m_0}$) is almost constant, given by
\begin{equation}
  \label{eq:57}
  \bE^x_{outside} \approx 2\bE_0 (m_0 d)\:.
\end{equation}
The expression~(\ref{eq:54}) is the main result of this paper. It shows that
effect of anomaly inflow leads to the drastic change of the electromagnetic
fields on the brane. Namely, in the presence of magnetic field, the electric
field behave as if photon has become massive, with the mass $m_0$ dependent on
the magnetic field. The same is true for a point-like charge, placed in
magnetic field~(\ref{eq:58}). In particular, the electric field, created by an
infinite capacitor, appears outside its plates (eq.~(\ref{eq:57})). As we will
argue in our next paper~\cite{anomaly-exp}, this effect can be used as a
signature of extra dimensions.

\subsubsection{Perturbative computations of the initial stage}
\label{sec:old-part}

Although we have already obtained a non-perturbative final state, we would
still like to see how this solution is formed. Therefore let us come back to
the dynamical problem. Although the naive perturbation theory did not work,
the parameter $\kappa_0$ is very small, and it is natural to assume that if we
will choose a more appropriate zeroth approximation, we can describe
time-dependent effects of initial stage of anomaly inflow
perturbatively.\footnote{This is similar in spirit to the quasi-classical
  expansion in quantum mechanics, where there is a non-perturbative in $\hbar$
  part of the wave-function $e^{\frac{i S}{\hbar}}$ plus perturbative series
  in $\hbar$ in a pre-exponential term.} %
The results of the previous section explain why the naive perturbation theory
of Section~\ref{sec:perturb} did not work. Essentially, this is due to the
fact that the solution in the bulk is non-perturbative in $\kappa_0$.
Therefore, we will try to find a dynamic solution (linear in time), based on
the non-perturbative zeroth approximation, derived from the static
solution~(\ref{eq:51}).

At first sight solution~(\ref{eq:51}) differs drastically from electric fields
of Section~\ref{sec:perturb}.  For example, the component $E_x$ and a charge
density distribution are symmetric with respect to an inversion $x\to -x$,
while the correction $E_x^\1$ to an electric field is anti-symmetric. This is
due to the fact that the full (time-dependent) Maxwell equations do not
possess such symmetry (unlike the equations for the static case). Thus, for
example, the point particle can have a (time-dependent) dipole moment at
initial stage of anomaly inflow (see below, Section~\ref{sec:dipole-moment}),
while such a dipole moment will be absent in the static
solution~(\ref{eq:58}). Additionally, static solution behaves as
$e^{-\kappa_0\Delta(z)}$ for $|z|$ outside the region~(\ref{eq:53}) around the
brane, which is very different from the $z$ dependence of the fields $E_x^\0$
and $E_z^\0$ in Section~\ref{sec:perturb}, which had a power law decay
(see~\cite{gauge-fields}).  However, we can still develop a perturbative (in
time) expansions for electric fields.
As a zeroth approximation we will take functions $E_x^\0$ and $E_z^\0$, whose
profiles in $z$ direction is $\chi_0(z)$ and $\p z \chi_0(z)$ correspondingly
(that is they go to zero exponentially fast for $|z|\gg \frac1{\M}$), unlike
functions $E_x^\0$ and $E_z^\0$ in the Section~\ref{sec:perturb}).  Namely, we
express $E_z^\0$ and $E_x^\0$ via a function $\Phi^\0(x,z)$ such that $E_z^\0
= \p z \Phi^\0(x,z)$ and $E_x^\0= \p x \Phi^\0(x,z)$. The function
$\Phi^\0(x,z)$ will be chosen in the form~(\ref{eq:51}):
\begin{equation}
  \label{eq:59}
  \Phi^\0(x,z)=\phi^\0(x)\chi_0(z)\quad\text{where}\quad \phi^\0(x) =
  -\frac{\bE_0}2\Bigl( |x+d|-|x-d|\Bigr)\:, 
\end{equation}
where $\bE_0$ is a value of a \emph{4-dimensional electric field} between the
plates of the capacitor.  We stress that $\Phi^\0(x,z)$ is just an auxiliary
function, which is not related to the component $\Phi^\0$ of the gauge field.
The initial field $\phi^\0(x)$ is created by an infinite capacitor in $3+1$
dimensions.

The reason for such a choice is the following. The relaxation time in the bulk
is determined by the mass of the fermions there (fermionic mass gap).  This
mass is much larger than any other scale in our theory, therefore one can
think that the bulk fermions form an incompressible fluid and have a
relaxation time $\tau_\psi$ much faster than any time scale in the theory.
Therefore the profile in $z$ direction settles over the time of turning on of
a magnetic fields. As the magnetic field $H^x$ is already taken at its
stationary value in our problem, we also take static profile in $z$-direction
for the electric fields.  On the contrary, the relaxation in $x$-direction is
of the order $\tau_x \sim m_0^{-1}$, which is incomparably smaller than the
afore-mentioned scale. The difference of these two scales justifies an
ansatz~(\ref{eq:59}). We will find corrections to the electric
field~(\ref{eq:59}) for the times $\tau_\psi \lll t \ll \tau_x$.  The
resulting expression has $z$ dependence of the form
$\frac{\chi_0(z)}{\Delta(z)}$ (see Appendix~\ref{app:correct} for details).
Thus, it will decay in the same region, as a full solution and a problems of
the perturbation theory of the Section~\ref{sec:perturb} are removed.

%
In the linear stage the result of the anomaly inflow is the following.  The
anomalous electric density appears between the plates of the capacitor:
\begin{equation}
  \label{eq:185}
  \rho_\anom(t,x)=  -6\,\kappa_0  \bE_0 \bH_0 t,\quad |x|<d\:,
\end{equation}
where $\bH_0$ is the value of magnetic field $H^\0_x$ in the center of the
solenoid.  This means that the \emph{capacitor acquires anomalous electric
  charge, linearly changing with time}. This implies an existence of the
stationary 4-dimensional current $\bj^x(x)$, given by
\begin{equation}
  \label{eq:186}
  \bj^x(x) = 6\kappa_0 t \bH_0\Phi^\0(x,0)\:,
\end{equation}
(where $\Phi^\0$ is given by~(\ref{eq:59})).  This picture has an apparent
contradiction with causality, as a non-zero current $\bj^x$ instantaneously
appears at spatial infinity at $t=0$. This is due to our choice of initial
conditions: the magnetic field $H_x^\0$ was turned on instantaneously in the
whole space. In reality there is a transitional period, depending on the speed
with which we are turning on magnetic field, and the current will appear at
infinity only when magnetic field will reach it.

We see that in accord with our qualitative expectations, electric charge is
initially accumulating in the region of space where initial $\vec E\cdot \vec
H\neq 0$. Such a charge of course creates an electric field outside the plates
of the capacitor. For $|x|>d$  and $r\ll R$ the electric
field outside of the capacitor behaves as
\begin{equation}
  \label{eq:187}
   \bE^\1_x= -12\kappa_0 e^2\, d \, \bE_0 \bH_0 \sign(x)t
   ,\quad |x|>d\:.
\end{equation}
Comparing correction $\bE^\1_x$ with $\bE_0$, we see that the linear stage of
anomaly inflow is valid for the time $t\ll \tau_{linear}$ where
\begin{equation}
  \label{eq:188}
   \tau_{linear} \sim \frac1{12\kappa_0 e^2\,d\, \bH_0} =\frac1{m_0 (m_0 d)}\:.
\end{equation}

\subsection{Anomalous field of elementary particles and dipole moment}
\label{sec:dipole-moment}

We discussed in Section~\ref{sec:observ-inflow} the simplest possible way to
create configuration of electro-magnetic field with parallel $E$ and $H$ and
therefore observe effects of anomaly inflow.  Actually, the electro-magnetic
field created by any charged particles with spin has non-zero $\vec E\cdot
\vec H$.  As we will presently see, anomaly inflow in this case leads to an
appearance of an anomalous dipole moment of the particle.  To estimate this
effect, we start from the quasi-classical expressions for electric and
magnetic fields, created by a particle with an electric charge $e$ and
magnetic moment $\mu=\frac em$
\begin{equation}
  \label{eq:77}
  \vec E = e\frac{\vec r}{r^3};\quad \vec H = \frac{3(\vec \mu\cdot \vec
    r)\,\vec r  - r^2\vec \mu}{r^5}
  \Rightarrow    \vec E\cdot \vec H= \frac{2(\vec\mu\cdot \vec r)e}{r^6}\:.
\end{equation}
Thus, in the region of the space where expression~(\ref{eq:77}) is non-zero,
inflow current (analogous to that of Section~\ref{sec:observ-inflow}) creates
non-zero anomalous charge density $\rho$. Such density will be positive in one
half of the ball, surrounding a particle, and negative in another one. As a
result, the total charge of the particle does not change.  However due to the
inflow any particle acquires an anomalous \emph{electric dipole
  moment}:
\begin{equation}
  \label{eq:78}
  d_\anom \sim e{(\rho \,r_c^3)}\, {r_c}
\end{equation}
(here $r_c$ is a Compton radius of the particle, given by $r_c \sim \frac 1m$
in our system of units).  Charge density $\rho$ is given by~(\ref{eq:65}) with
electric and magnetic fields estimated by $E\sim \frac{e}{r_c^2}=e\, m^2$ and
$H\sim \frac{e}{m \,r_c^3} = e m^2$.  Substituting all these values into
eq.~(\ref{eq:78}) we get:
\begin{equation}
  \label{eq:79}
 \boxed{ d_\anom\sim e^3  \kappa_0\,t}\:,
\end{equation}
i.e. as a consequence of anomaly inflow a particle acquires a dipole moment,
which has the absolute value growing with time. Notice, however, that this
result was obtained in perturbation theory as a linear in time approximation
(Section~\ref{sec:old-part}). It is valid for $t\ll \tau_{particle}\sim \frac
1{m_0}$ where $m_0$ is given by an analog of formula~(\ref{eq:49}) for $H\sim
e m^2$:
\begin{equation}
  \label{eq:98}
  m_0 \sim m  \sqrt{\kappa_0 e^3} \:.
\end{equation}
For times, much bigger that this characteristic time, field configuration
around the particle approaches static solution. As it has already been
discussed (see the beginning of the Section~\ref{sec:old-part}), static
equations are symmetric with respect to inversion $\vec x \to - \vec x$,
therefore, there cannot be any dipole moment for $t\gg \tau_{particle}$.
Electric field configuration will be significantly modified from the usual
Coulomb to Yukawa form (similar to~(\ref{eq:58})) at the distances larger than
$m_0^{-1}$, with $m_0$ given by~(\ref{eq:98}). This means that the electric
charge of the particle \emph{gets completely screened} and the total amount of
an anomalous charge which appeared on the brane is equal to the charge of the
particle.


\section{The Standard Model with anomaly inflow}
\label{sec:anom-sm}

In this section we apply the logic of Section~\ref{sec:qed} to the Standard
Model.  Indeed, if the Standard Model fields are localized on a brane in a
5-dimensional world, there is no apparent reason to expect a separate anomaly
cancellation for them.
Let us try to see what kind of new effects we should expect for the Standard
Model if 4-dimensional anomaly cancellation condition is not imposed and there
is an inflow from extra dimensions. We will see that such effects do exist and
discuss experimental restrictions for corresponding parameters.  For
simplicity we consider the electroweak $SU(2)\times U(1)$ theory with only one
generation of fermions (the addition of extra generations does not change the
analysis). The action for the $SU(2)$ and $U(1)$ gauge fields is similar to
that in eq.~(\ref{eq:1}). The warp-factor $\Delta(x^4)$ is such that there is
a zero mode, localized on the brane. We also add a Higgs $\phi$ field, which
is an $SU(2)$ doublet. As in Section~\ref{sec:qed5d} its mass $m_\phi^2(x^4)$
is negative at $x^4=0$ and tends to the positive constant in the bulk as
$|x^4|\to \infty$ (Figure~\ref{fig:higgs}). Therefore it has a non-trivial
profile $\phi(x^4)$, with $\phi(0)\neq 0$ and $\phi(x^4)\to 0$ at $x^4\to\pm
\infty$.  So the $SU(2)\times U(1)$ symmetry is broken down to
$U(1)_{\textsc{em}}$ on the brane and restored far from it.  Therefore the
vector boson acquires mass $\m(x^4)$ which becomes zero at large
$|x^4|$.\footnote{%
  Notice, that one could not simply have chosen $\m(x^4)=\const$, as such a
  constant would have removed the zero mode from the spectrum.  Indeed, in
  this paper we consider such warp-factors $\Delta(x^4)$ that there is a
  mass-gap $\M$ between the zero mode and the modes in the bulk. Therefore no
  state with the mass $\m \ll \M$ can exist away from the brane.}

We also add fermions in the bulk, charged with respect to $SU(2)\times U(1)$.
Their zero modes are localized on the brane with mass gap $m_\psi$ much larger
than that of the gauge fields.

\subsection{Charge difference of electron and proton and anomalies}
\label{sec:anom-charges}

\begin{figure}[t]
  \centering \includegraphics[scale=.5]{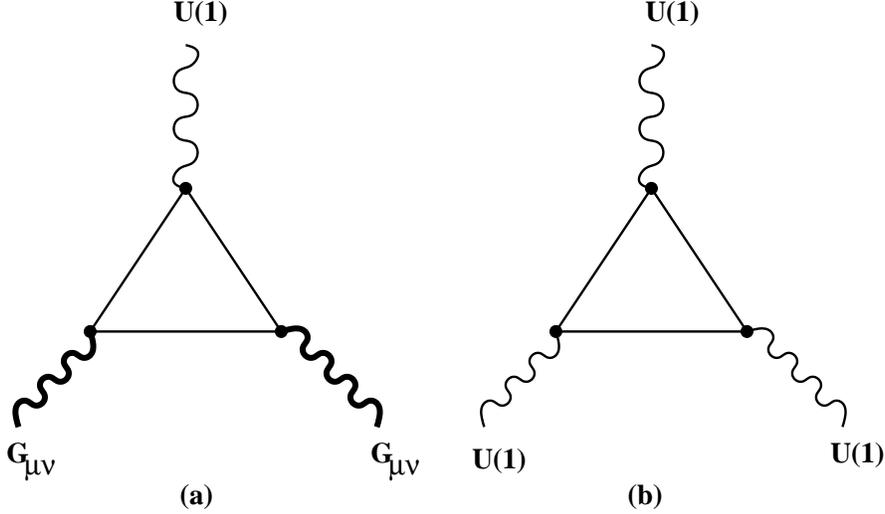}
  \caption{Anomalous diagrams, describing the non-conservation of
    the $U(1)$ current in the presence of non-trivial $SU(2)$ background
    \textbf{(a)}; or in a background of two $U(1)$ fields~\textbf{(b)}.}
  \label{fig:anomaly}
\end{figure}

We consider the extension of the Standard Model in which the $U(1)_Y$
hyper-current $j_\Y^\mu$ becomes anomalous
\begin{equation}
  \label{eq:92}
  \p\mu j_\Y^\mu =  \frac{\Tr[Y^3]}{16\pi^2}
  \epsilon^{\mu\nu\lambda\rho}\CF_{\mu\nu}\CF_{\lambda\rho}
  +\frac{\Tr[Y_L]}{16\pi^2}   \epsilon^{\mu\nu\lambda\rho}\Tr
 G_{\mu\nu} G_{\lambda\rho}\:.
\end{equation}
Here $\CF_{\mu\nu}$ is a $U(1)_\Y$ field strength of the $U(1)$ field $B_\mu$
(hyper-photon): $\CF_{\mu\nu} = \p\mu B_\nu - \p\nu B_\mu$;
$G^\alpha_{\mu\nu}$ is an $SU(2)$ non-Abelian field strength, $g'$ and $g$ are
$U(1)_\Y$ and $SU(2)$ coupling constants correspondingly.  The first term
in~\eqref{eq:92} comes from the diagram Fig.~\ref{fig:anomaly}b and is
proportional to the sum of cubes of hypercharges of \emph{all} particles. The
hypercharges in the Standard Model are chosen in such a way that
$\Tr[Y^3]=\Tr[Y_L] = \Tr[Y]=0$~\cite{sm-anom}. We choose a one-parameter
extensions of this model, in which
$\frac16\Tr[Y^3]=-\frac12\Tr[Y_L]=\kappa_0$. The second term comes from the
diagram Fig.~\ref{fig:anomaly}a, with one $U(1)_\Y$ and two $SU(2)$
vertices.\footnote{Note, that no anomalies arise for QCD!} %
Along with anomaly~\eqref{eq:92} there is also a non-conservation of $SU(2)$
current in the background $U(1)$ and $SU(2)$ fields (its coefficient is
proportional to the same diagram as the second term in~\eqref{eq:92}):
\begin{equation}
  \label{eq:94}
    D^\mu j_\mu^\alpha =  \frac{\Tr[Y_L]}{8\pi^2}
  \epsilon^{\mu\nu\lambda\rho}G^\alpha_{\mu\nu} F_{\lambda\rho}\:.
\end{equation}
Recall that $U(1)_\Y$ and $SU(2)$ fields, entering~\eqref{eq:92}
and~\eqref{eq:94}, are the fields \emph{above} the electroweak symmetry
breaking scale.  At lower energies it is convenient to re-express these
anomalies in terms of of electro-magnetic field $\gamma_\mu$ and
\emph{neutral} field $Z_\mu$, which can be obtained from the $U(1)$ and
$SU(2)$ fields $ B_\mu$ and $A_\mu^3$ via the rotation $ \gamma_\mu = B_\mu
\cos^2\theta_w + A^3_\mu\sin^2\theta_w$ and $Z_\mu = (A^3_\mu -
B_\mu)\cos\theta_w \sin\theta_w$.\footnote{These relations differs from the
  usual ``textbook'' ones ($\gamma_\mu = B_\mu \cos\theta_w +
  A^3_\mu\sin\theta_w$, etc.) because we use different normalization of the
  action for the gauge fields -- the one with the coupling constant in front
  of the kinetic term.} 
The electro-magnetic current $j_{\gamma}^\mu$ is given by $j^\mu_\Y+j_3^\mu$
and \emph{neutral current} $j_{\z}^\mu = -\cot\theta_w j_\Y^\mu+\tan\theta_w
j^{\mu}_3$, where $j^{\mu}_3$ is the 3rd component of the $SU(2)$ triplet
$j_\alpha^\mu$, $\alpha=1,2,3$.  Using~(\ref{eq:92}) and~(\ref{eq:94}) one can
easily see that \emph{(i)} electro-magnetic current is conserved in the
arbitrary background of electro-magnetic fields (as one expected, because the
electrodynamics remains vector-like in our model); \emph{(ii)} there is an
anomalous $\gamma\gamma Z$ \emph{coupling}
\begin{equation}
  \label{eq:96}
  \p\mu j^\mu_\z = -\frac{2\,\,N_f\,\kappa_0}{\pi^2\cos\theta_w\sin\theta_w}
  \vec E_\gamma\cdot  \vec H_\gamma\:,
\end{equation}
which implies the non-conservation of the neutral current in the parallel
electric and magnetic fields ($N_f$ is the number of generations);
\emph{(iii)} another important consequence of the presence of $\gamma\gamma Z$
coupling is the non-conservation of the electro-magnetic current in the mixed
electro-magnetic and $Z$-backgrounds:
\begin{equation}
  \label{eq:97}
  \p\mu j_\gamma^\mu =  -\frac{4\,N_f\,\kappa_0}{\pi^2\cos\theta_w
    \sin\theta_w} (\vec E_\gamma \cdot  \vec H_\z + \vec E_\z \cdot  \vec
  H_\gamma) \:.
\end{equation}
As we will see this leads to the effects similar to those, described in
Sections~\ref{sec:observ-inflow}--\ref{sec:dipole-moment}.

\subsection{Static electric field in a capacitor in a magnetic field}
\label{sec:change-E}

Consider again the setup of Section~\ref{sec:observ-inflow}: place a capacitor
in the strong magnetic field $\vec H$, such that electric field $\vec E$,
created by the capacitor is in the same direction $x^1$.  Our choice of
hypercharges implies the non-conservation of $j^\mu_\z$ current~\eqref{eq:96}.
This also means that there exist Chern-Simons terms in the bulk action, which
ensure an anomaly inflow, canceling anomaly~(\ref{eq:96}).  Indeed,
Chern-Simons term~(\ref{eq:3}), originally written in terms of the hypercharge
field $B_\mu$ and $SU(2)$ field $A^a_\mu$, can also be re-expressed in terms
of electro-magnetic and \textsc{Z}-field and it creates a inflow of
\textsc{Z}-current in the parallel electric and magnetic fields.  This creates
anomalous density of \textsc{Z} charge on the brane. This distribution of
\textsc{Z} charge creates an anomalous \textsc{Z} field in the 5th direction
and as a result inflow of \emph{electro-magnetic} current to cancel an
anomaly~(\ref{eq:97}). Such an inflow creates anomalous distribution of
electric charge on the brane and modifies electric field inside and outside
the capacitor. Similarly to the case of electrodynamics
(Section~\ref{sec:qed}) we can find this static configuration of an electric
field. It will be given by the same expression as in
Section~\ref{sec:final-state} (see Appendix~\ref{sec:ew-final-state-compute}
for details). As a result, distribution of an electric field inside and
outside the capacitor is given by expressions~(\ref{eq:55})--(\ref{eq:57}),
with $m_0$ proportional to the magnetic field $\bH_0$ and anomaly parameter
$\kappa_0$:
\begin{equation}
  \label{eq:237}
  \boxed{m_0^2 = \frac{N_f e^2
    \bH_0}{\pi^2\cos\theta_w\sin\theta_w}\,\kappa_0 }\:.
\end{equation}
Notice, that this result does not depend on the mass scale of the extra
dimension and would be true even for Planck scale $\M$.  It also does not
depend on the mass of the \textsc{Z} boson $\m_\z$.  This feature, as well as
the result~(\ref{eq:237}) itself, is valid only assuming that $\m_\z\ll\M$.
This can be understood as follows: anomalous density of \textsc{Z} charge
creates electric \textsc{Z} field in the 5th direction. This field is, of
course, decaying on the scales larger than $1/{\m_z}$.  However the inflow
comes from the region~(\ref{eq:53}) which is much smaller than $1/{\m_z}$.  As
a result, in the leading order in ${\m_\z}/{\M}$ mass of the \textsc{Z} boson
does not modify the effect.  For details see
Appendix~\ref{sec:ew-final-state-compute}.


\subsection{Anomalous dipole moment}
\label{sec:proton}

The consequence of anomaly~\eqref{eq:98} is the appearance of anomalous dipole
moment of the particle (c.f.  Section~\ref{sec:dipole-moment}). Namely, a
background with non-zero $\vec E_\gamma \cdot \vec H_\z+\vec E_\z \cdot \vec
H_\gamma$ can be created around any particle which has a spin and also
electric and $Z$-charges.  Unlike the electro-magnetic case, $Z$-field,
created by the particle, is short-ranged.
In the physically interesting case of a particle, lighter than $Z$-boson (e.g.
proton or electron), anomaly inflow is concentrated in the region smaller then
Compton wave length of the particle and, therefore, it can not be treated
quasi-classically. Therefore we will only give the order of magnitude estimate
here. Compared to the analysis of Section~\ref{sec:dipole-moment} the result
is suppressed by $(\frac{m}{\m_\z})^2$ :
\begin{equation}
  \label{eq:104}
   d_\anom \sim \frac{N_fe^2 \kappa_0}{\cos\theta_w\sin\theta_w}
  \left(\frac{m}{\m_\z}\right)^2\: t \:.
\end{equation}
Again, this result does not depend on the parameters of the extra dimension,
but in this case it depends on the mass of the particle and on $\m_\z$.

\section{Discussion}
\label{sec:disc}

In this paper we analyzed a brane-world scenario, when fields on the
brane possess an anomaly, canceled by inflow from the bulk. We stress
that in such setup there is no reason to require cancellation of
anomaly separately on the brane. We show in a simple model of
electrodynamics that inflow results in the concrete observational
effects, even in the situation when the real escaping of the matter
from the brane is not possible.

This logic can also be applied to the Standard Model (SM), if one
considers it as an effective theory of some fundamental theory in
higher-dimensional space-time. Anomaly cancellation condition which
is usually assumed in SM becomes an unjustified fine-tuning and one
should study the theory without it. Easing this condition leads (in
the simplest case) to the one free parameter -- electric charge
difference of electron and proton.  This is one more legitimate
phenomenological parameter imposed onto the SM from the point of view
of extra-dimensions.  We show that with this parameter (called
$\kappa$ in sections~\ref{sec:anom-sm}--\ref{sec:proton}) being
non-zero, the phenomenology of the Standard Model is modified in such
a way as to admit an anomaly.  It is interesting to note that this
modification does not affect electrodynamics sector of the SM: even
for different charges of electron and proton it remains
non-anomalous.

The presence of anomaly inflow exhibits itself on a brane via a subtle
mechanism, leading to the dynamical screening of an electric field in the
presence of a parallel magnetic field, e.g the change of an electric field in
a capacitor, placed in a magnetic field.  Similarly, anomaly inflow leads to
the screening and change with time of the electric charge of any elementary
particle with electric charge and spin.  the inflow, before the screened
static solution is formed, such a particle acquires also anomalous dipole
moment, which, however, disappears again in the final state, where the
electric field is parity symmetric, but screened.

These effects would not be present if the theory were purely 4-dimensional,
with anomaly canceled by addition of new (chiral) particles at higher
energies. Thus the described effects can in principle serve as a signature of
extra dimensions.  Of course, there can exist other low-energy (as well as
high-energy) effects, through which anomaly inflow exhibits itself. For
example, one can show that photon, propagating in the magnetic field, would
becomes massive, depending on its polarization.  We plan to study other
signatures of anomaly inflow elsewhere.

\paragraph{Radiative corrections.} 

\begin{figure}[t]
  \centering \includegraphics[scale=.6]{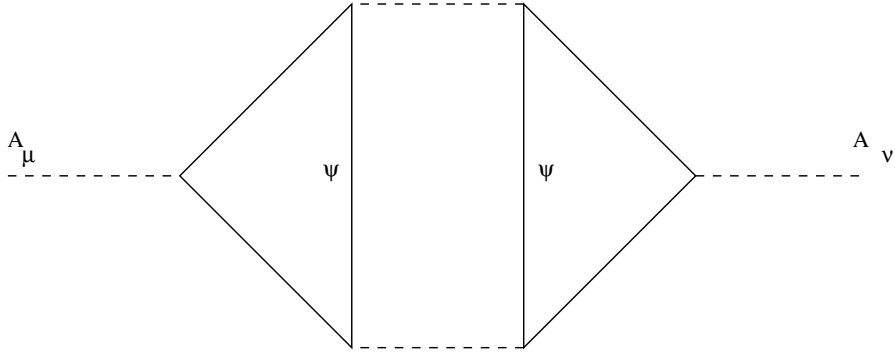}
  \caption{A diagram potentially generating the photon mass in electroweak
  theory.}
  \label{fig:twoCS}
\end{figure}

The analysis in this paper was conducted at the level of classical equations
of motion.  The natural question arises: to which extent this theory can be
considered weakly coupled and radiative corrections can be neglected? A most
sensible quantity seems to be the mass of the photon, which is constrained
strongly by a number of different experiments. A simple power counting applied
to a diagram, shown on Fig.~\ref{fig:twoCS}, containing two triangles in
the {\it anomalous four-dimensional} electroweak theory with a non-standard
hypercharge choice, gives an order of magnitude of the photon mass $m_\gamma
\sim e^3 \Lambda_c\kappa_0$. Then, for an ultraviolet cutoff of the order of
$1$ TeV one finds a very strong constraint on an anomaly coefficient
$\kappa_0$. If the same consideration were true for our case, when the
anomalous electroweak theory is coming from anomaly-free theory in 5
dimensions, the classical effects described in this paper would be subleading
and non-observable.  However, this is not the case because of the following
reason.  The same diagram, considered now in full 5-dimensional theory, where
both massive and zero modes run in loops, does not lead to generation of the
photon mass simply because the theory is free from anomalies and is
gauge-invariant. On the language of the effective field theory, the insertion
of triangular diagrams will not produce any irregularities, as any
contribution coming from the Chern-Simons vertex is precisely canceled by the
same process with an insertion of a D'Hoker-Farhi vertex.

\paragraph{Effective theory.}

Our main result, given by eqs.~(\ref{eq:54}), (\ref{eq:237}), does not depend
on the mass scale of the 5th dimension $\M$ and on the Higgs VEV. So, the
effect stays even when these parameters are sent to infinity, in other words,
when the only low energy particle residing in the spectrum is the
photon.\footnote{Strictly speaking, in this limit there are no particles which
  can create background magnetic or electric fields.} This can also be seen
from eqs.~(\ref{eq:3}) and~(\ref{eq:20}), which do not contain any information
about the heavy particles which generated them and are not suppressed by any
cut-off.  This may seemingly contradict to the usual logic behind effective
field theories which is based on the conjecture (often known as ``decoupling
theorem'' and proven for quite a general class of four-dimensional theories in
\cite{Appelquist}) that effects of massive fields on the {\it renormalizable}
low-energy effective action only exhibit themselves in renormalization of
charges and fields, while all additional interaction are suppressed by some
positive power of $\left(\frac{E}{\Lambda_{c}}\right)$ (characteristic energy
of processes over cut-off $\Lambda_c$).  It is known, however, that the
``decoupling theorem'' does not always hold.  In particular, it is not true in
theories, leading to Chern-Simons~\cite{redlich} or D'Hoker-Farhi~\cite{df}
interactions.  In the latter case the low energy theory is {\it
  non-renormalizable}, which is true for our case as well. So, in addition to
the Chern-Simons and the D'Hoker-Farhi term in the effective theory one must
introduce infinitely many other terms and counterterms to remove divergences.
To determine them one needs the knowledge of the fundamental theory. The
construction of this type of theory goes beyond the scope of the present work.

The low energy description of our theory contains 5-dimensional terms (see
e.g.  action~(\ref{eq:2})). It would be interesting to construct a low-energy
effective action entirely in terms of 4-dimensional fields.  Naively, one
could try to integrate over the massive modes of the field $A_a(x,z)$ in the
action~(\ref{eq:2}). However, as this work demonstrates, the Chern-Simons term
in~(\ref{eq:2}) cannot be treated perturbatively, therefore it is not clear
how one can perform such an integration.

\paragraph{Fine-tuning of anomaly mismatch.}
The idea of realization of the Standard Model on a brane is very popular and
wide-spread nowadays. As we showed in this paper, any such model should either
produce a mechanism, which prohibits the anomaly on the brane (due to symmetry
reasons or dynamics) or incorporate it into the model.  At the same time, it
is known experimentally that (if non-zero) the charge difference between
electron and proton is extremely small.  As a result, one is presented with a
novel type of a fine-tuned parameter and should try to find a reason for its
existence.  As it is usually the case, this fine-tuning can tell us something
new about the physics beyond the Standard Model.

Certainly, the answers to the questions of anomalous Standard Model
and fine-tuning of its anomaly coefficient will depend on various
ingredients of the brane-world models: the higher-dimensional theory,
the details of localization of the fields, etc.  Therefore it is
important to consider semi-realistic brane-worlds, where at least
some sector of the Standard model (including gauge fields) is
localized. A simple example of such sort was proposed
in~\cite{Shap.qed}, where the $U(1)$ theory was localized on the
vortex in six dimensions. In that model the fermionic content on the
brane could have anomalous chiral couplings (while the theory in the
bulk was always anomaly free). This theory can be considered as a
particular realization of the scenario of section~\ref{sec:qed} and
effects, similar to those, described in
section~\ref{sec:observ-inflow}, should be present there. One should
notice also that its anomaly coefficient can be of the order of
unity, as it should on general grounds. More realistic models should
include non-Abelian fields localized which is rather non-trivial.  In
all these models it is important to analyze the question of anomaly
(non)cancellation on the brane and the reason of the fine-tuning of
corresponding parameters.

In string theory, being a theory with extra-dimensions, this question
also arises.  A popular approach there is to obtain the Standard
Model fields on an intersection of various branes (see e.g. recent
paper~\cite{Antoniadis}). On the other hand, there are many examples
in string/M theory with branes when world volume theory is anomalous
and there is an inflow~(see e.g.~\cite{I-brane}).  Again, there
should be a special reason to have so exact anomaly cancellation for
the Standard Model realized in this way. Anomaly analysis is usually
very instructive in string theory and a string theoretical solution
of this problem would be very interesting.

One should also apply this logic to any extension of the Standard
Model (e.g. to the Minimal Supersymmetric Standard Model (MSSM))
appearing in a brane-world setup. In this case there could be more
free parameters which may allow for some other phenomenological
effects (e.g.  non-zero electric charge of neutralinos).

\subsection*{Acknowledgments}

We would like to thank I.~Antoniadis, X.~Bekaert, V.~Cheianov, T.~Damour,
S.~Dubovsky, J.~Harvey, S.~Khlebnikov, M.~Libanov, J.-H.~Park,
S.~Randjbar-Daemi, S.~Shadchin for useful comments.  The work A.B. and M.S.
was supported by the Swiss Science Foundation.  O.R. would like to acknowledge
a partial support of the European Research Training Network contract 005104
"ForcesUniverse" and also warm hospitality of Ecole Polytechnique F\'ed\'erale
de Lausanne where a part of this work was done.  A.B. would like to
acknowledge the hospitality of Institut des Hautes \'Etudes Scientifiques.

\appendix

\section{Perturbation theory in $\kappa_0$}
\label{app:linear}

To compute initial, linear in time, change of electromagnetic fields on the
brane, one should specify values of the fields at $t=0$. As discussed at the
beginning of Section~\ref{sec:perturb}, one should specify not only values of
electromagnetic field (to be discussed below, in
Sections~\ref{app:naive}--\ref{app:correct}), but also initial values of gauge
potential. In this paper we consider the following case.
\begin{equation}
  \label{eq:23}
  A_\theta(r,z)\biggr|_{t=0}\hskip -1ex = \int^r_0\hskip -.2em d r\, r
  H^x(r,z);\qquad A^a\biggr|_{t=0}\hskip -1ex = 0\quad \forall\: a\neq
  \theta  \:. 
\end{equation}
Values of the time derivative at $t=0$ can then be  determined, knowing field
strengths and using eq.~(\ref{eq:18}).  There are two non-zero derivatives:
\begin{equation}
  \label{eq:24}
  \p t A_x\biggr|_{t=0} = E_x(x,z);\qquad  \p t A_z\biggr|_{t=0} = E_z(x,z)\:.
\end{equation}
Our initial conditions mean that at the initial moment the longitudinal
component of the field is not excited. Indeed, at $t=0$ our field $A^a$ obeys:
$A^0=0$ and $\div \vec A=0$ (we denote by $\div \vec A \equiv \p x A^x + \p r
A^r + \frac 1r \p\theta A^\theta + \p z A^z$), which means that it is
transversal. Notice, that the condition~(\ref{eq:24}) implies that
transversality of $A^a$ will not hold for $t>0$.
For these initial conditions the only non-zero component of the D'Hoker-Farhi
current is
\begin{equation}
  \label{eq:30}
 j^r\biggr|_{t=0}\hskip -1ex = -6\kappa_0  A_\theta(r,z) E_x(x,z);\qquad  j^0_\df=j^x_\df=j^\theta_\df\biggr|_{t=0}\hskip -1ex=0\:.
\end{equation}

\subsection{Naive perturbation theory}
\label{app:naive}

First, we attempt to solve the Maxwell equations~(\ref{eq:6})--(\ref{eq:7}) by
perturbation theory in $\kappa_0$. In addition to the initial
condition~(\ref{eq:30}), we take initial electric fields to satisfy Gauss
constraints in the theory with $\kappa_0=0$ (see the discussion in the
beginning of Section~\ref{sec:perturb}):
\begin{equation}
  \label{eq:31}
  \p z \Bigl(\Delta(z) E^z_\0\Bigr) + \Delta(z) \p x E^x_\0 = \sigma_0\delta(z)
  \Bigl(\delta(x+d) -\delta(x-d)\Bigr)\:.
\end{equation}
We will mark such fields by the symbol $^\0$. Our choice in particular imply
that $J^0_\cs|_{t=0} = 0$.  Although the zero mode of the gauge field has
constant profile in fifth direction, the solution of~(\ref{eq:31}) gives
$E^x_\0(x,z)$ and $E^z_\0(x,z)$, both decaying at $|z|\to\infty$. The rate at
which these fields decay depends on the warp-factor $\Delta(z)$.  A solenoid
creates non-zero field $H^x (r,z)$ as well as $F^{\theta z}(r,z)$, again, both
decaying at infinity in $z$ (see~\cite{gauge-fields} for details).  According
to the definition of the inflow current~(\ref{eq:8}) such electric and
magnetic fields generate the inflow current from the fifth direction:
\begin{equation}
  \label{eq:32}
  J^z_\cs = 6 \kappa(z) E_x^\0(x,z) H_x^\0(r,z)\:.
\end{equation}
This current flows onto the brane from both sides in the $z$ direction.  Two
other non-zero components of the Chern-Simons current are
\begin{equation}
  \label{eq:33}
  J^x_\cs = - 6 \kappa(z) E_z^\0(x,z) H_x^\0(r,z) , \quad J^r_\cs =
  -6\kappa(z) F_{\theta z}^\0(r,z) E_x^\0(x,z)\:. 
\end{equation}
For our configuration of electric and magnetic fields the D'Hoker-Farhi
current~(\ref{eq:20}) satisfies the following property:
\begin{equation}
  \label{eq:34}
  \p tj^0_\df + \p x j^x_\df =\p rj^r_\df + \frac 1r\p\theta j^\theta_\df =
  \frac 12\p\mu j^\mu_\df  = -6 \kappa_0 E_x^\0(x,z) H_x^\0(r,z) \delta(z)\:.
\end{equation}
This property suggests that the dynamics of the theory may be separated into
two parts: dynamics in the directions $x,z$ and in $r,\theta$. We will see
below that this is indeed the case at the initial stage of the process.  For
the initial conditions of Section~\ref{sec:perturb}, the anomaly of the
D'Hoker-Farhi current is split equally between $\p r j^r_\df$ and $\p t
j^0_\df$, both being equal to the right hand side of eq.~(\ref{eq:34}). In
particular
\begin{equation}
  \label{eq:86}
  \p t j^0_\df(x^\mu)\biggr|_{t=0}\hskip -1ex  = -6\kappa_0 E_x^\0(x,z)
  H_x^\0(r,z) \delta(z) \:,
\end{equation}
i.e. we see that an electric charge is accumulating on the brane between the
plates of the capacitor. 

At initial stage of anomaly inflow, only electric fields appears in the left
hand side of the Maxwell equations~(\ref{eq:6})--~(\ref{eq:7}). The reason for
that is the following. As we have shown, initially the total charge density
grows linearly in time~(\ref{eq:86}).  Therefore, it creates electric field,
also growing linearly in time. As a consequence of the Bianchi identities ($\p
t F_{ij}^\1 = \p i E_j^\1 -\p j E_i^\1$) all the magnetic components
$F_{ij}^\1, F_{iz}^\1=\CO(t^2)$. Thus, if we keep only terms at most linear in
time in these equations, we obtain:
\begin{align}
  &\hphantom{ - }\frac1{\e^2}\p z \Bigl(\Delta(z)E_z^\1\Bigr) +
  \frac1{\e^2}\Delta(z)\bigl(\p x E_x^\1 +\p r E_r^\1\bigr) =
  - j^0_\df(x^\mu,z)\delta(z)\label{eq:36}\:,\\
  & - \frac1{\e^2}\Delta(z)\p t E_x^\1 = 6\kappa(z) E_z^\0(x,z) H^\0_x(r,z)
  \label{eq:37}\:,\\
  &- \frac1{\e^2}\Delta(z)\p t E_r^\1 = -j^r_\df(x^\mu,z)\delta(z) +6\kappa(z)
  E_x^\0(x,z) F^\0_{\theta z}(r,z)\label{eq:38}\:,\\
  &-\frac1{\e^2}\Delta(z)\p t E_z^\1= -6\kappa(z)
  E_x^\0(x,z)H^\0_x(r,z)\label{eq:39}
\end{align}
  All the gauge fields in the left
hand side are of the order $\CO(\kappa_0)$, while those in the right hand side
are of the zeroth order in $\kappa_0$ and are explicitly multiplied by
$\kappa_0$.  Eq.~(\ref{eq:37}) gives
\begin{equation}
\label{eq:40}
    E_x^\1(x^a) = -6\e^2\frac{\kappa(z)}{\Delta(z)}
    E_z^\0(x,z)H^\0_x(r,z)t+\CO(t^2) \:,
\end{equation}
and similarly from eq.~(\ref{eq:39})
\begin{equation}
  \label{eq:41}
  E_z^\1(x^a) = 6\e^2\frac{\kappa(z)}{  \Delta(z) }
  E_x^\0(x,z)H^\0_x(r,z)t+\CO(t^2) \:.
\end{equation}
We see that \emph{for any} $\kappa_0$ the correction $E^\1_x$ will become
bigger than $E^\0_x$ for large $|z|$, as $\Delta(z)\to 0$ for $|z|\to \infty$.

Notice, that solution~\eqref{eq:40} is a \emph{5-dimensional} electric field.
To discuss the observational consequences, we should switch to the
4-dimensional fields, defined as in~(\ref{eq:12}), namely
\begin{equation}
\label{eq:42}
  \frac1{e^2} \vec\bE(x^\mu) = \frac1{\e^2}\int^{\infty}_{-\infty}
  dz\,\Delta(z) \vec E(x^\mu,z) \:.
\end{equation}
Four-dimensional electric field $\bE^\1_x$ is given by
\begin{equation}
  \label{eq:43}
  \bE^\1_x(x^\mu) = -12e^2 \kappa_0 t \int^{\infty}_{0} dz
  E_z^\0(x,z)H^\0_x(r,z) \:.
\end{equation}
Integration in~\eqref{eq:43} can be done explicitly in the region $r\ll R$. As
shown in \cite{gauge-fields} $H_x^\0(r,z)$ is a constant as a function of $z$
for $\M^{-1}\lesssim |z|\ll \M R^2$.  Notice, that even in case of
$H_x^\0=\const$ integral~(\ref{eq:43}) is convergent.  Indeed, for any
warp-factor $\Delta(z)$, decaying at infinity, $E_z$-component of an electric
field decays at infinity as $|z|^{1+\epsilon}$ (see~\cite{gauge-fields} for
details).  We will be using this approximation through this section. The error
from approximation of $H^x(r,z)$ by its value at the origin
$H_x^\0(r{=}0,z{=}0)=\bH_0$ can be estimated as $\int^\infty_{\M R^2} dz\,
E_z^\0 H_x^\0 \sim\CO(\frac 1{\M R})\to 0$ (see~\cite{gauge-fields} for
details). In case of the warp $\Delta(z)=e^{-2\M |z|}$ the
integral~(\ref{eq:43}) is saturated in the region $|z|\sim \M x^2$.

The solution of this Section has an apparent problem: for any $\kappa_0$, e.g.
$E_x^\1$ becomes bigger than $E_x^\0$ for $|z|$ large enough. See the
discussion in the Section~\ref{sec:perturb}.

\subsection{Perturbative solution for the capacitor}
\label{app:correct}

To compute perturbatively in $\kappa_0$ an initial changes of electric field
of a capacitor, we should choose a zeroth approximation different from that of
the previous Appendix. Namely, we express $E_z^\0$ and $E_x^\0$ via an
auxiliary function $\Phi^\0(x,z)$ such that $E_z^\0 = \p z \Phi^\0(x,z)$ and
$E_x^\0= \p x \Phi^\0(x,z)$ with $\Phi^\0(x,z)$ given by~(\ref{eq:51}):
\begin{equation}
  \label{eq:110}
  \Phi^\0(x,z)=\phi^\0(x)\chi_0(z)\quad\text{where}\quad \phi^\0(x) = -\frac{\bE_0}2\Bigl( |x+d|-|x-d|\Bigr)\:,
\end{equation}
where $\bE_0$ is a value of a \emph{4-dimensional electric field} between the
plates of the capacitor.  The initial field $\phi^\0(x)$ satisfies
4-dimensional Poisson equation: $-\p x^2\phi^\0(x) =
e^2\sigma_0\Bigl(\delta(x+d) -\delta(x-d)\Bigr)$.

Unlike the case of Section~\ref{sec:perturb}, the \emph{initial} Chern-Simons
charge density $J^0_\cs$ is not equal to zero (as one can see by substituting
$E^\0_x$ and $E^\0_z$ into the Gauss constraint). This implies that component
$F_{xz}^\0$ is nonzero (it can be read off the right hand side of the Gauss
constraint~(\ref{eq:22}) using the definition~(\ref{eq:8}) (see
also~(\ref{eq:158})):
\begin{equation}
  \label{eq:60}
    F^{xz}_\0 = \left(\Bigl( \p x^2\phi^\0(x) -m_0^2 \phi^\0(x) \Bigr)\Delta(z)
    +\frac{\e^4\kappa_0^2 \bigl(H^\0_x\bigr)^2}{\Delta(z)}\phi^\0(x)\right)
  \frac{\chi_0(z)\sign(z)}{\e^2\kappa_0H^\0_x} \:,
\end{equation}
where $m_0^2$ is given by~(\ref{eq:49}).  However, its explicit form will not
be needed for the analysis below. The first correction to the field $E_x$ is
given by
\begin{equation}
  \label{eq:61}
  E^\1_x(x,z,t) =  -6\e^2\frac{\kappa(z)}{\Delta(z)}
    E_z^\0(x,z)H^\0_x(r,z)t +\frac{\p z \Bigl(\Delta(z) F_{xz}^\0\Bigr)}{\Delta(z)}t\:,
\end{equation}
(recall, that our initial conditions for the D'Hoker-Farhi currents are those
of eq.~(\ref{eq:30}), and therefore $j^x_\df = 0$). Similarly
\begin{equation}
  \label{eq:62}
  E_z^\1(x^a) = 6\e^2\frac{\kappa(z)}{  \Delta(z) }
  E_x^\0(x,z)H^\0_x(r,z)t - \frac{\p x \Bigl(\Delta(z)
    F_{xz}^\0\Bigr)}{\Delta(z)}t\:. 
\end{equation}
Notice that according to the definition~(\ref{eq:12}) the \emph{4-dimensional}
electric field, $\bE^\1_x$ is determined by only the first term
in~(\ref{eq:61}):
\begin{equation}
  \label{eq:63}
  \bE^\1_x(t,x,z) =  -12e^2 \kappa_0 t \int^{\infty}_{0}\hs dz\,
  E_z^\0(x,z)H^\0_x(r,z) \:.
\end{equation}
The difference of equation~(\ref{eq:63}) with that of~(\ref{eq:43}) in the
Section~\ref{sec:perturb} is in the fact that here the integral over $z$ in
effectively restricted to the region $z\lesssim\frac1{\M}\log\frac{\M}{\e^2
  \kappa_0 \bH_0}$. In this region one can substitute $H_x^\0(r,z)$ with its
value $\bH_0$ in the center of the solenoid. As a result one gets:
\begin{equation}
  \label{eq:64}
  \bE^\1_x(t,x) = 12e^2 \kappa_0 t \bH_0\int^{x}_0\hs dx\,
  E_x^\0(x,z\,{=}\,0)\:. 
\end{equation}

One can check that in the linear in $\kappa_0$ order the energy is conserved.
Indeed, in this order the change of energy (in the full 5-dimensional theory)
is given by
\begin{equation}
  \label{eq:74}
  \Delta\CE_{5d} = \frac12\int d^3 x \int dz\, \Delta(z)\bigl( E_x^\0E_x^\1 +
  E_z^\0 
  E_z^\1\bigr)\:.
\end{equation}
Substituting the solutions~(\ref{eq:61})--(\ref{eq:62}) we see that the
integrand of~(\ref{eq:74}) is equal to zero.  From the point of view of the
4-dimensional observer, the energy is defined as
\begin{equation}
  \label{eq:75}
  \CE_{4d}=\frac12\int\hs d^3 x\, \vec \bE^2 \:,
\end{equation}
and therefore its change  in the linear in $\kappa_0$ order is
\begin{equation}
  \label{eq:76}
    \Delta\CE_{4d} = \int\hs d^3 x\, \bE_x^\0\bE_x^\1 \:.
\end{equation}
Because the $\bE^\0(x)$ is an even function of $x$, while $\bE^\1(x)$ is an
odd one, this integral is equal to zero. Thus although there is an inflow of
Chern-Simons currents to the brane, the energy on the brane does not change.
It simply gets redistributed in the space, as field appear outside the
capacitor and diminishes inside.

According to the definition~(\ref{eq:110}) $E_x^\0(x,z{=}0) = \p x
\phi^\0(x)\chi_0(0) = \bE_0 \Theta(d^2 -x^2)$,
where the function $\Theta(d^2 -x^2)$ equals to 1 for $|x|<d$ and zero
otherwise and $\chi_0(0)=1+\CO(\kappa_0)$ (see
Appendix~\ref{sec:final-state-compute}).  As a result we see that \emph{the
  inflow current creates a non-zero charge density on the brane in the region
  where (4-dimensional) $\vec E\cdot \vec H\neq 0$} (in our case -- between
the plates of the capacitor):\footnote{In computing $\div\vec \bE^\1$ one
  should take into account that there is a contribution, coming from the term
  $\p r \bE^\1_r$.  As mentioned after the eq.~(\ref{eq:34}), divergence of
  the D'Hoker-Farhi current is equally split between $t,x$ and $r,\theta$
  components. Therefore, the coefficient in front of the integral
  in~(\ref{eq:65}) is twice as little as that in eq.~(\ref{eq:64}). This can
  be easily verified by directly computing $E^\1_r$ from eq.~(\ref{eq:38}) and
  substituting it together
  with~(\ref{eq:64}) into~(\ref{eq:65}).} %
\begin{equation}
  \label{eq:65}
   \rho_\anom(x^\mu) = \frac1{e^2}\div\vec \bE^\1 = 6 \kappa_0 t \int^\infty_0
   \hskip -.25em dz \,\p x E_z^\0(x,z) 
   H_x^\0(r,z)\approx 
   -6\, t\,\kappa_0  \bE_0 \bH_0 \,\Theta(d^2 -x^2)\:.
\end{equation}
Thus the capacitor accumulates over time an additional electric charge
$Q_{\anom} = 2d\,S\,\rho_\anom $ (where area $S=\pi R^2$).  This means that
from a 4-dimensional point of view there is an electric current flowing from
infinity. Indeed, according to the definition~(\ref{eq:14}) the 4-dimensional
current in the direction $x$ is given by
\begin{equation}
  \label{eq:66}
  \bj^x(x) = \int^{\infty}_{-\infty}dz J^x_\cs(x,z)\:.
\end{equation}
(both $j^x_\df$ and $F_{xz}^\1$ are equal to zero in the linear in time
approximation). As we will now see, this current $\bj^x$ is non-zero as
$|x|\to\infty$.  Using definition~(\ref{eq:33}) we get:
\begin{equation}
  \label{eq:67}
  \bj^x(x) = 6\kappa_0 \bH_0\Phi^\0(x,0)\:.
\end{equation}
where $\Phi^\0(x,0)\equiv \phi^\0(x)\chi_0(0)$.  If anomalous electric charge
$Q_\anom$ changes with time, this means that $S\int\hs dx\, \p x\bj^x \neq 0$
or equivalently that $\bj^x(+\infty)-\bj^x(-\infty)\neq 0$ (we have
substituted integral over the $(r,\theta)$ plane with the area $S$ of the
plates of the capacitor).  Using eq.~(\ref{eq:110}) one can see that indeed
$\Phi^\0(\pm \infty,0) = \pm\bE_0 d$. This gives
\begin{equation}
  \label{eq:68}
  \frac{d Q_\anom}{d t} =
  -\bj^x(x)\biggr|^{+\infty}_{-\infty} S = -12\kappa_0  \bH_0\bE_0 S d\:.
\end{equation}

The result~(\ref{eq:68}) can be obtained from the conservation of the
5-dimensional current.  Namely, one can see that there is a flow of the
Chern-Simons current from infinity. In the approximation $H^x(r,z)=\bH_0$ we
should also take $F^{\theta z}=0$ ($F^{\theta z}$ has the same $z$-dependence
as $\p z H_x^\0$) and therefore we can neglect radial Chern-Simons current. As
a result the analysis becomes effectively $2+1$ dimensional in coordinates
$(t,x,z)$.

Consider a box $|x|\le L_x$, $|z|\le L_z$, where both $L_x,L_z\to \infty$.
The total amount of Chern-Simons current, inflowing through the boundaries of
this box is
\begin{equation}
  \label{eq:69}
  \frac {d Q_\cs}{d t} \equiv \int^{L_x}_{-L_x}\hs dx
  \Bigl(J^z_\cs(x,L_z)-J^z_\cs(x,-L_z)\Bigr)+\int^{L_z}_{-L_z}\hs dz
  \Bigl(J_\cs^x(L_x,z)-J_\cs^x(-L_x,z)\Bigr) \:.
\end{equation}
From eqs.~(\ref{eq:32})--(\ref{eq:33}) it follows that the first integral
equals to zero:
\begin{equation}
  \label{eq:70}
  \int^{L_x}_{-L_x}\hs dx
  \Bigl(J^z_\cs(x,L_z)-J^z_\cs(x,-L_z)\Bigr) = 12\kappa_0 \bH_0
  \int^{L_x}_{-L_x}\hs dx\, \p x\Phi^\0(x, L_z) = 0\:,
\end{equation}
as $\Phi^\0(x,L_z)\to 0$ for $L_z\to \infty$ (see~(\ref{eq:110})). At the same
time the second integral in~(\ref{eq:69}) is finite:
\begin{equation}
  \label{eq:71}
\int^{L_z}_{-L_z}\hs dz
  \Bigl(J_\cs^x(L_x,z)-J_\cs^x(-L_x,z)\Bigr)=6\bH_0
  \int^{L_z}_{-L_z}\hs dz\,\kappa(z) \p z \Phi^\0(L_x,z)=12\kappa_0
  \bH_0\Phi^\0(L_x,0) \:,
\end{equation}
as $\Phi^\0(x,z=0)\to \pm \bE_0 d$ for $x\to \pm\infty$. As a result we again
recover~(\ref{eq:68}).\footnote{This is not surprising of course.
  Eq.~(\ref{eq:9}) together with the initial conditions $j^x_\df=0$ and the
  fact that the total charge density is equal to $j^0_\df$ implies that
  $Q_\cs$ in~(\ref{eq:69}) is equal to $Q_\anom$.}

The anomalous charge $Q_\anom$ creates an additional electric field
eq.~(\ref{eq:43}).  That is we see that due to the anomaly
inflow an electric field appears outside of the capacitor. In the region close
to the plates, (i.e. $|x|\sim d$ and $r\ll R$) this field is constant in space
and grows linearly in time according to the law
\begin{equation}
  \label{eq:72}
  \bE^\1_x\approx -12\kappa_0 e^2\, t\, d \,   \bE_0 \bH_0 \:.
\end{equation}
Inside the capacitor $\bE^\1_x$ changes linearly from its value~(\ref{eq:72})
on one plate of the capacitor to the opposite of it on another.
We see that the electric field, created due to the anomaly inflow depends on
the anomaly coefficient $\kappa_0$ and \emph{does not depend} on the
parameters of the extra dimension. From eq.~(\ref{eq:72}) we can determine the
characteristic time during which the linear approximation is valid. It is
given by
\begin{equation}
  \label{eq:73}
  \tau \sim \frac1{12\kappa_0 e^2\,d\, \bH_0} =\frac1{m_0 (m_0 d)}\:,
\end{equation}
and our solution~(\ref{eq:40}) is valid for $t\ll \tau$.

\section{Static solution of Maxwell equations}
\label{sec:final-state-compute}

Consider the system of Maxwell equations, describing a \emph{static} solution
in $4+1$ dimensions:
\begin{align}
  \p z \Bigl(\Delta(z) E^z \Bigr) + \Delta(z)\Bigl(\p x E^x + \p r E^r\Bigr)
  &=\e^2\Bigl(
  q(x)\delta(z) + j^0_{\df} + J^0_\cs\Bigr)\:,\label{eq:153}\\
  \p z \Bigl(\Delta(z) F^{xz} \Bigr) +\frac {\Delta(z)}r \p r \Bigl(r
  F^{xr}\Bigr) &= \e^2\Bigl(j^x_\df + J^x_\cs\Bigr)\:,\label{eq:154}\\
  \p z \Bigl(\Delta(z) F^{rz} \Bigr) + \Delta(z) \p x
  F^{rx}&= \e^2\Bigl(j^r_\df + J^r_\cs\Bigr)\:,\label{eq:155}\\
  \p z \Bigl(\Delta(z) F^{\theta z} \Bigr) +\Delta(z) \p x F^{\theta x}+\frac
  {\Delta(z)}r \p r \Bigl(r F^{ \theta r}\Bigr) &= \e^2\Bigl(j^\theta_\df +
  J^\theta_\cs\Bigr)\label{eq:156}\:,\\
  \Delta(z)\Bigl(\p x F^{xz} +\frac 1r \p r \bigl(r F^{rz}\bigr)\Bigr) &=
  -\e^2 J^z_\cs\label{eq:157}\:.
\end{align}
(We will be mostly interested in the case when $q(x) = \sigma_0
\bigl(\delta(x+d)-\delta(x-d)\bigr)$ or $q(x) = q_0 \delta(\vec x)$).  We are
searching for the axially-symmetric static solution, therefore all the
derivatives with respect to time and angle $\theta$ were put to zero in
eqs.~(\ref{eq:153})--(\ref{eq:157}).

The components of Chern-Simons current~(\ref{eq:8}) in the cylindrical
coordinates $x,r,\theta$ are equal to
\begin{align}
  J_\cs^0 &= \frac{6\kappa(z)}{r} \Bigl(F_{xr} F_{\theta z} - F_{x\theta}
  F_{rz} +
  F_{xz}F_{r\theta}\Bigr)\:,\label{eq:158}\\
  J_\cs^x &= -\frac{6\kappa(z)}{r} \Bigl(E_r F_{\theta z} + E_z
  F_{r\theta}\Bigr)\:,\label{eq:159}\\
  J_\cs^r & = \frac{6\kappa(z)}{r} \Bigl(E_x F_{\theta z} + E_z
  F_{x\theta}\Bigr)\:,\label{eq:160}\\
  J_\cs^\theta & = -\frac{6\kappa(z)}{r} \Bigl(E_x F_{r z} - E_r F_{xz} + E_z
  F_{xr}
  \Bigr)\:,\label{eq:161}\\
  J_\cs^z & =\frac{6\kappa(z)}{r}\Bigl(E_x F_{r\theta} - E_r
  F_{x\theta}\Bigr)\:.\label{eq:162}
\end{align}
For static solution we can describe the electric field in terms of the
electro-static potential $\Phi$: $E_i = -\p i \Phi$, $E_z = -\p z \Phi$. In this
case one can rewrite expression $J^\mu_\cs + j^\mu_\df$ in the following way:
\begin{equation}
  \label{eq:163}
  J^a_\cs + j^a_\df = \frac34 \p b \Bigl(\kappa(z) \Phi
  F_{cd}\epsilon^{abcd0}\Bigr) ,\quad a \neq 0\:.
\end{equation}
As a result we solve the system~(\ref{eq:153})--(\ref{eq:157}) by the following
ansatz. We express $F_{xz}$, $F_{rz}$, $F_{xr}$ via $\Phi$ and $F_{\theta i}$,
using eqs.~(\ref{eq:154}), (\ref{eq:155}), (\ref{eq:157}):
\begin{equation}
  \label{eq:164}
  F^{xz} = \frac{6\e^2\kappa(z)\Phi F_{r\theta}}{r\Delta(z)},\quad F^{xr} =
  \frac{6\e^2\kappa(z)\Phi F_{\theta z}}{r\Delta(z)},\quad F^{rz} =
  \frac{6\e^2\kappa(z)\Phi F_{x\theta}}{r\Delta(z)} \:.
\end{equation}
Then the Gauss constraint~(\ref{eq:153}) can be re-written as
\begin{equation}
\label{eq:165}
  \p z \Bigl(\Delta(z) \p z \Phi\Bigr) + \Delta(z) \nabla^2 \Phi = \frac{36\e^4
    \kappa_0^2}{r^2 \Delta(z)}\Bigl( F_{r\theta}^2 + F_{x\theta}^2 + F_{\theta
      z}^2\Bigr) \Phi+ \e^2 q(x)\delta(z)+j^0_\df\:,
\end{equation}
where $\nabla^2$ is a 3-dimensional Laplacian in the coordinates $x,r,\theta$.
The equation~(\ref{eq:156}) becomes (we introduce gauge potential $A_\theta$:)
\begin{equation}
  \label{eq:166}
  \p z \Bigl(\Delta(z) \p z A_\theta\Bigr) + \Delta(z) \nabla^2 A_\theta =
  \frac{36\e^4 \kappa_0^2 \Phi}{r^2 \Delta(z)}\Bigl(\p x A_\theta \,\p x \Phi  +
  \p z A_\theta\p z \Phi + \p r A_\theta \p r \Phi\Bigr)\:.
\end{equation}
We will find a solution of the eqs.~(\ref{eq:165})--(\ref{eq:166}) as follows.
First we will put right hand side of eq.~(\ref{eq:166}) to zero. Then there is
a solution $F_{r\theta} = r H^x$, $H^x=\const=\bH_0$, $F_{x\theta} = F_{\theta
  z}=0$.  In this case one could easily find an explicit solution of
eq.~(\ref{eq:165}) -- see eqs.~(\ref{eq:167})--(\ref{eq:176}) below. After that
we will show that corrections to the found solutions, which are due to the
non-zero right hand side of are of the order $\kappa_0^2$ and thus can be
neglected.

Under assumptions $F_{r\theta} = r \bH_0$, $\bH_0=\const$, $F_{x\theta} =
F_{\theta z}=0$ Gauss law becomes identical to that of the $2+1$ dimensional
case.  After the Fourier transform in $x,r,\theta$-directions: ($\nabla\to i
\vec p$) and a substitution $\Phi=\frac{\psi_p(z)}{\sqrt{\Delta(z)}}$
equation~(\ref{eq:165}) becomes\footnote{We have neglected a D'Hoker-Farhi
  charge density $j^0_\df$ in eq.~(\ref{eq:9}) as compared to~(\ref{eq:165}).
  It accounts for the perturbative corrections of the order $\CO(\kappa_0)$.}
\begin{equation}
\label{eq:167}
  \psi_p''(z) +\psi_p(z) \Bigl[W'(z) - W^2(z) - p^2\Bigr] =
  \frac{36\e^4\kappa_0^2 \bH_0^2}{\Delta^2(z)} \psi_p(z)+\e^2 \tilde q(\vec p)
  \delta(z) \:,
\end{equation}
where $W(z) \equiv -\frac{\Delta'(z)}{2\Delta(z)} $ and $\tilde q(\vec p)$ is
a Fourier transform of $q(x)$. For our main example -- the warp-factor
$\Delta(z) = e^{-2\M |z|}$, we have $W(z) = \M\sign z$. Let us re-write
eq.~(\ref{eq:167}) as
\begin{equation}
  \label{eq:168}
  (\hat H + p^2) \psi_p(z) = -\e^2 q_0 \delta(z)\:,
\end{equation}
where operator $\hat H$ does not depend on $p$:
\begin{equation}
\label{eq:169}
  \hat H = -\p z^2 + \Bigl(36\e^4\kappa_0^2 \bH_0^2\, e^{4\M|z|} + \M^2 -2
  \M\delta(z)\Bigr) \:.
\end{equation}

\paragraph{Spectrum of $\hat H$:}
\label{sec:spectrum}

Let us denote eigen-functions of~(\ref{eq:169}) and their eigen-values by
$\psi_n(z)$ and $m_n^2$: $\hat H\psi_n(z) = m_n^2\psi_n(z)$.  Introduce new
\emph{dimensionless} variable $\xi \equiv \frac{3\e^2\kappa_0\bH_0}{\M} e^{2 M
  |z|}$ and denote by $\nu_n^2 = \frac{\M^2-m_n^2}{4\M^2}$. Then the
eigen-value problem for operator~(\ref{eq:169}) reduces to
\begin{equation}
  \label{eq:170}
  \xi^2\p\xi^2 \psi_n(\xi) +\xi\p\xi \psi_n(\xi) -(\nu_n^2+\xi^2)\psi_n(\xi) =
  0 \:.
\end{equation}
This is the standard form of the equation for the Bessel functions of the
second kind.  A solution of this equation, which is regular at $\xi\to\infty$
(i.e. $z\to\infty$) is given by $K_{\nu_n}(\xi)$, which behaves at infinity as
$K_\nu(\xi) \sim \frac{e^{-\xi}}{\sqrt\xi}$. As a result we see that the
solution of eq.~(\ref{eq:170}) is given by
\begin{equation}
  \label{eq:171}
  \psi_n(z) = \,K_{\nu_n}\left( \xi_0\,
    e^{2 M |z|}\right) \quad \text{where}\quad\nu_n =
  \tfrac12\sqrt{1-\tfrac{m_n^2}{\M^2}}\quad \text{and}\quad\boxed{\xi_0\equiv
  \tfrac{3\e^2\kappa_0\bH_0}{\M}  }\:.
\end{equation}
Eigen-values $m_n^2$ are determined from the gluing condition at the origin,
following from the presence of $\delta(z)$ in the operator~(\ref{eq:169})
$\bigl[\psi_n'(z)\bigr]\Bigr|_{z=0} = -2\M\psi_n(0)$:
\begin{equation}
  \label{eq:172}
  2\xi_0
  K_{\nu_n}'(\xi_0) +
  K_{\nu_n}(\xi_0) =0\:,
\end{equation}
where $\xi_0$ was defined in~(\ref{eq:171}) and $\xi_0\ll 1$.  Using the
properties of the Bessel function (see e.g.~\cite{gradshteyn}) one can show
that there is a single solution of this equation for real $\nu_n$
(corresponding to the smallest eigen-value ($m_0 < \M$) and infinitely many
solutions for imaginary $\nu_n$'s (corresponding to all the other eigen-values
$m_n>\M$).

The easiest way to solve eq.~(\ref{eq:172}) is by iteration around $\nu_0 =
\frac12$. As a result one finds that\footnote{One can show for an arbitrary
  function $\Delta(z)$, for which a zero mode exists, that operator $\hat H$
  has the lowest eigen-value $m_0^2 \sim \kappa_0 \e^2 \M$.}
\begin{equation}
  \label{eq:173}
  m_0^2 = 4\xi_0 \M^2(1 -\xi_0\log \xi_0) + \CO(\xi_0^2)\:.
\end{equation}
The wave-function of the lowest level is given by
\begin{equation}
\label{eq:174}
  \psi_0(z) = 
  c_0\,K_{\nu_0}\left(\xi_0 e^{2 M |z|}\right)\quad
  \text{where} \quad \nu_0 \approx \frac12-\xi_0\:,
\end{equation}
where $K_\nu(u)$ is a modified Bessel function of the second kind and $c_0$ is
a normalization constant, given by $c_0 =
\sqrt{\frac{2\M\xi_0}{\pi}}+\CO(\xi_0^{3/2}\log\xi_0)$.

The Green's function of the operator~(\ref{eq:168}) is given by the standard
expression
\begin{equation}
\label{eq:175}
  G(z) = \sum_{n=0}^\infty \frac{\psi_n(z)\bar\psi_n(0)}{m_n^2+p^2}\:.
\end{equation}
The solution of eq.~(\ref{eq:165}) can be easily found for any charge
distribution $q(x)$. In case at hand (capacitor with the infinite plates) one
gets:
\begin{equation}
  \label{eq:176}
  \Phi(x,z) = -\e^2 {\sigma_0} \sum_{n=0}^\infty \left( e^{-m_n |x-d|} - 
  e^{-m_n |x+d|}\right) \frac{\psi_n(z)\bar\psi_n(0)}{2m_n\sqrt{\Delta(z)}}  \:.
\end{equation}
We will also often use the function $\chi_0(z) \equiv
\frac{\psi_0(z)\bar\psi_0(0)}{\sqrt{\Delta(z)}}$. It is given by:
\begin{equation}
  \label{eq:177}
  \chi_0(z)=\frac{2 \M\xi_0}{\pi\sqrt{\Delta(z)}} 
  K_{{\nu_0}}(\xi_0 \,e^{2\M|z|})K_{{\nu_0}}(\xi_0)=
  \sqrt{\frac{2\xi_0}{\pi}}\frac{\M}{\Delta(z)}K_{\frac 12 - \xi_0}\left(\tfrac{\xi_0}{\Delta(z)}\right)+\CO(\xi_0\log\xi_0)\:,
\end{equation}
and has  the following important properties:
\begin{equation}
  \label{eq:105}
  \chi_0(z{=}0) = \M\:,
\end{equation}
and
\begin{equation}
  \label{eq:178}
  \int^\infty_{-\infty}\hs dz \,\Delta(z) \chi_0(z) = 1+\CO(\xi_0\log\xi_0)\:.
\end{equation}

\section{Static solution in electroweak theory}
\label{sec:ew-final-state-compute}

Consider the system of two couple $U(1)$ fields: a massless (electro-magnetic)
one, whose gauge potential and field strength we will denote by $A^a$ and
$F^{ab}$ and \emph{massive} ($\textsc{Z}$ field), with gauge potential $\CA^a$
and field strength $\CF^{ab}$. These fields interact via Chern-Simons terms:
\begin{equation}
  \label{eq:235}
  S_{int} = \int\kappa_\ggz(z) AF\CF + \kappa_\gzz A\CF\CF + \kappa_\zzz
  \CA\CF\CF \:.
\end{equation}
where (compare~(\ref{eq:96}))
\begin{equation}
\small 
 \label{eq:84}
  \kappa_\ggz =-\frac{N_f\kappa_0}{4\pi^2\cos\theta_w
    \sin\theta_w},\;
  \kappa_\gzz =-\frac{N_f\kappa_0 (\cos^3\theta_w-3\sin^3\theta_w)}{\pi^2\cos^3\theta_w \sin^3\theta_w},\;
  \kappa_\zzz =\frac{3N_f\kappa_0\cos2\theta_w}{16\pi^2\cos^3\theta_w
    \sin\theta_w} 
\end{equation}
Fermions interact with both electro-magnetic and \textsc{Z} fields.  According
to eqs.~(\ref{eq:92}),~(\ref{eq:94}) both electro-magnetic current
$j^\mu_\gamma$ and \textsc{Z}-current $j^\mu_\z$ are not conserved:
\begin{equation}
  \label{eq:217}
  \p\mu j^\mu_\z = \kappa'_\ggz(z)
  \epsilon^{\mu\nu\lambda\rho}F_{\mu\nu}F_{\lambda\rho} + 2\kappa'_\gzz(z)
  \epsilon^{\mu\nu\lambda\rho}\CF_{\mu\nu}F_{\lambda\rho} + \kappa'_\zzz(z)
  \epsilon^{\mu\nu\lambda\rho}\CF_{\mu\nu}\CF_{\lambda\rho}\:,
\end{equation}
and
\begin{equation}
  \label{eq:226}
  \p\mu j^\mu_\gamma =
  2\kappa'_\ggz(z)\epsilon^{\mu\nu\lambda\rho}\CF_{\mu\nu}F_{\lambda\rho} 
   + \kappa'_\gzz(z) \epsilon^{\mu\nu\lambda\rho}\CF_{\mu\nu}\CF_{\lambda\rho}\:.
\end{equation}
This means that there are two types of inflow currents: \textsc{Z}-current
\begin{equation}
  \label{eq:218}
  \frac{\delta S_\eff}{\delta \CA^a}\equiv \CJ^a_{\cs,\z}=
   \kappa_\ggz(z) \epsilon^{abcde}   F_{bc}  F_{de} + 
   \kappa_\gzz(z) \epsilon^{abcde} \CF_{bc}  F_{de} +
   \kappa_\zzz(z) \epsilon^{abcde} \CF_{bc}\CF_{de}\:,
\end{equation}
and electro-magnetic Chern-Simons currents:
\begin{equation}
  \label{eq:219}
   \frac{\delta S_\eff}{\delta A^a}\equiv J^a_{\cs,\gamma}=
   \kappa_\ggz(z) \epsilon^{abcde}   F_{bc}\CF_{de} + 
   \kappa_\gzz(z) \epsilon^{abcde} \CF_{bc}\CF_{de}\:.
\end{equation}
The expression for the D'Hoker-Farhi currents $j^\mu_{\df,\z}$ and
$j^\mu_{\df,\gamma}$ are similar to that of eq.~(\ref{eq:20}) with an obvious
modifications following from eqs.~(\ref{eq:217})--(\ref{eq:226}).

Consider the system of Maxwell equations, describing a \emph{static} solution
in $4+1$ dimension. For static solution we can describe the electric field in
terms of the electro-static potential $\Phi$: $E_i = -\p i \Phi$, $E_z = -\p z
\Phi$. We will denote electro-static potential for the electro-magnetic and
\textsc{Z}-field as $\Phi_\gamma$ and $\Phi_\z$ correspondingly. Then we obtain:
\begin{align}
  \p z \Bigl(\Delta(z) \p z \Phi_\gamma \Bigr) + \Delta(z)\nabla^2\Phi_\gamma
  &=-\e^2\Bigl(q(x)\delta(z) + j^0_{\df} + J^0_{\cs,\gamma}\Bigr)\:,\label{eq:179}\\
  \p z \Bigl(\Delta(z) F^{xz} \Bigr) +\frac {\Delta(z)}r \p r \Bigl(r
  F^{xr}\Bigr) &= \e^2\Bigl(j^x_\df + J^x_{\cs,\gamma}\Bigr)\:,\label{eq:180}\\
  \p z \Bigl(\Delta(z) F^{rz} \Bigr) + \Delta(z) \p x
  F^{rx}&= \e^2\Bigl(j^r_\df + J^r_{\cs,\gamma}\Bigr)\:,\label{eq:181}\\
  \p z \Bigl(\Delta(z) F^{\theta z} \Bigr) +\Delta(z) \p x F^{\theta x}+\frac
  {\Delta(z)}r \p r \Bigl(r F^{ \theta r}\Bigr) &= \e^2\Bigl(j^\theta_\df +
  J^\theta_{\cs,\gamma}\Bigr)\:,\label{eq:182}\\
  \Delta(z)\Bigl(\p x F^{xz} +\frac 1r \p r \bigl(r F^{rz}\bigr)\Bigr) &=
  -\e^2 J^z_{\cs,\gamma}\:,\label{eq:183}
\end{align}
where $\nabla^2$ is a 3-dimensional Laplacian in the coordinates $x,r,\theta$.
We will be mostly interested in the case when $q(x) = \sigma_0
\bigl(\delta(x+d)-\delta(x-d)\bigr)$ or $q(x) = q_0 \delta(\vec x)$.  
For the \textsc{Z}-field we have similar system of equations:
\begin{align}
  \p z \Bigl(\Delta(z)\Phi_\z \Bigr) + \Delta(z)\nabla^2\Phi_\z -\e^2\m^2(z)
  \Phi_\z &=-\e^2\Bigl(q_\z(x)\delta(z)+j^0_{\df,\z} +
  \CJ^0_{\cs,\z}\Bigr)\:,\label{eq:220}\\
  \p z \Bigl(\Delta(z) \CF^{xz} \Bigr) +\frac {\Delta(z)}r \p r \Bigl(r
  \CF^{xr}\Bigr)-\e^2\m^2(z) \CA^x
  &= \e^2\Bigl(j^x_{\df,\z} + \CJ^x_{\cs,\z}\Bigr)\:,\label{eq:221}\\
  \p z \Bigl(\Delta(z) \CF^{rz} \Bigr) + \Delta(z) \p x \CF^{rx} -\e^2\m^2(z)
  \CA^r
  &= \e^2\Bigl(j^r_{\df,\z} + \CJ^r_{\cs,\z}\Bigr)\:,\label{eq:222}\\
  \p z \Bigl(\Delta(z) \CF^{\theta z} \Bigr) +\Delta(z) \p x \CF^{\theta
    x}+\frac {\Delta(z)}r \p r \Bigl(r \CF^{ \theta r}\Bigr)-\e^2\m^2(z)
  \CA^\theta &= \e^2\Bigl(j^\theta_{\df,\z} +
  \CJ^\theta_{\cs,\z}\Bigr)\:,\label{eq:223}\\
  \Delta(z)\Bigl(\p x \CF^{xz} +\frac 1r \p r \bigl(r
  \CF^{rz}\bigr)\Bigr)+\e^2\m^2(z) \CA^z &= -\e^2
  \CJ^z_{\cs,\z}\:.\label{eq:224}
\end{align}
We are searching for the axially-symmetric static solution, therefore all the
derivatives with respect to time and angle $\theta$ were put to zero in
eqs.~(\ref{eq:179})--(\ref{eq:224}).

To find a solution of this non-linear system of equations, let us recall the
electro-magnetic case first (Appendix~\ref{sec:final-state-compute}). We saw
there that of all the magnetic components $F_{ij}$ and $F_{iz}$ only
$F_{r\theta}$ was not suppressed by a power of anomaly parameter
$\kappa_0$.\footnote{As all parameters $\kappa_\ggz$, $\kappa_\gzz$ and
  $\kappa_\zzz$ are of the same order, we will use notation $\kappa_0$ when
  speaking about the order of magnitude value of various expressions.}
We will try to find a solution of the system~(\ref{eq:179})--(\ref{eq:224})
under a similar assumption: only terms, containing $F_{r\theta}$, give
dominating (non-perturbative in $\kappa_0$) contribution to these equations,
while all other terms account for small (suppressed by powers of $\kappa_0$)
corrections. In addition we take all components of the magnetic
\textsc{Z}-field to be much smaller than those of electro-magnetic field.
Then the system of equations~(\ref{eq:179})--(\ref{eq:224}) reduces to the
Gauss constraints for potentials $\Phi_\z$ and $\Phi_\gamma$, \emph{in which
  we have left only terms, proportional to $F_{r\theta}$}:
\begin{align}
  \label{eq:227}
  \p z \Bigl(\Delta(z) \p z \Phi_\gamma\Bigr) + \Delta(z) \nabla^2 \Phi_\gamma
  ={} & \frac{\e^2\kappa_\ggz(z)}{r} F_{r\theta}\CF_{xz}+\e^2 q(x)\delta(z)\:,
\end{align}
and
\begin{align}
  \label{eq:228}
  \p z \Bigl(\Delta(z) \p z \Phi_\z\Bigr) + \Delta(z) \nabla^2 \Phi_\z -
  \m^2(z)\Phi_\z &= \frac{\e^2\kappa_\gzz(z)}r F_{r\theta}\CF_{xz}
  +\frac{\e^2\kappa_\ggz(z)}r F_{r\theta}F_{xz} + \e^2 q_\z(x)\delta(z)\:.
\end{align}
Other relevant parts of the system~(\ref{eq:179})--(\ref{eq:224}) are
equations for the component $F_{xz}$:
\begin{align}
  \p z(\Delta(z) F_{xz}) &= \frac{\e^2}{r}\p z\Bigl(\kappa_\ggz(z) \Phi_\z
  F_{r\theta}
  \Bigr)\:,\label{eq:229}\\
  \p x(\Delta(z) F_{xz}) &= \frac{\e^2}{r}\p x\Bigl(\kappa_\ggz(z) \Phi_\z
  F_{r\theta} \Bigr)\:,
\end{align}
and $\CF_{xz}$:
\begin{align}
  \p z\Bigl(\Delta(z) \CF_{xz}\Bigr)-\m^2(z)\CA^x & = \frac{\e^2}r\p
  z\Bigl(\kappa_\ggz(z) \Phi_\gamma F_{r\theta} +\kappa_\gzz(z)\Phi_\z
  F_{r\theta} \Bigr)\:,\label{eq:230}\\
  \p x\Bigl(\Delta(z) \CF_{xz}\Bigr)+\m^2(z)\CA^z & = \frac{\e^2}r \p
  x\Bigl(\kappa_\ggz(z) \Phi_\gamma F_{r\theta} +\kappa_\gzz(z)\Phi_\z
  F_{r\theta} \Bigr)\:.\label{eq:231}
\end{align}
From eqs.~(\ref{eq:229}) we immediately see that
\begin{equation}
  \label{eq:225}
  F_{xz} = \frac{\e^2\kappa_{\ggz}(z)\Phi_\z F_{r\theta}}{r\Delta(z)}\:,
\end{equation}
similarly to the electro-magnetic case (compare eq.~(\ref{eq:164})).  Solution
of eq.~(\ref{eq:230}) depends on the $z$-dependence of the mass-term $\m(z)$.
As an example we will consider $\m^2(z) = \m_\z\delta(z)$ (localization scale
of the Higgs field is of the same order as that of the fermions on). In this
case, one can easily see that the solution of
eqs.~(\ref{eq:230})--(\ref{eq:231}) does not differ from the massless
case:\footnote{%
  Function $\CA^z$ is an odd function of $z$ and therefore $\delta(z)\CA^z=0$
  and mass term disappears from~(\ref{eq:231}). One could in principle add to
  the solution~(\ref{eq:225}) the term $\frac{m_\z \CA_x(x,0)\sign
    z}{\Delta(z)}$, however, this term grows as $\frac1{\Delta(z)}$ as
  $|z|\to\infty$ and therefore the energy of such a solution would be
  infinite. Thus, we conclude that $\CA_x(x,z{=}0)=0$, mass terms in
  eqs.~(\ref{eq:230})--(\ref{eq:231}) drop out and the solution for $\CF_{xz}$
  is given by~(\ref{eq:225}).}$^,$\footnote{The profile
  $\m^2(z)=\m_0\delta(z)$ is a limiting case, in which mass dependence drops
  out of equation~(\ref{eq:232}) and subsequent does not enter at all into
  Gauss constraint for an electric field. One can show that for arbitrary
  profiles $\m(z)$, spreading into the 5th direction, eq.~(\ref{eq:232})
  receives corrections due to mass. In particular the effect, described in
  this section disappears as $\m_\z\to \infty$.}
\begin{equation}
  \label{eq:232}
  \CF_{xz} = \frac{\e^2(\kappa_{\ggz}(z)\Phi_\gamma
    +\kappa_\gzz(z)\Phi_\z) F_{r\theta}}{r\Delta(z)} \:,
\end{equation}
and as a result we obtain the following system of equations:
\begin{align}
  \p z \Bigl(\Delta(z) \p z \Phi_\gamma\Bigr) + \Delta(z) \nabla^2 \Phi_\gamma
  & = \frac{\e^4\kappa_\ggz^2 \Phi_\gamma \bH_0^2}{\Delta(z)}+\e^2
  q(x)\delta(z)+ \frac{\e^2\kappa_\ggz\kappa_\gzz \Phi_\z\bH_0^2}{\Delta(z)}\:,
  \label{eq:233} \\
  \p z \Bigl(\Delta(z) \p z \Phi_\z\Bigr) + \Delta(z) \nabla^2 \Phi_\z -
  \m_\z\delta(z)\Phi_\z &= \frac{\e^4(\kappa_\ggz^2 +\kappa_\gzz^2)\Phi_\z
    \bH_0^2}{\Delta(z)} + \e^2 q_\z(x)\delta(z)\label{eq:234}\\
  & +\frac{\e^4\kappa_\ggz\kappa_\gzz \Phi_\gamma\bH_0^2}{\Delta(z)}\notag\:.
\end{align}
We will solve this system by assuming that the last terms of both
equations~(\ref{eq:233}) and~(\ref{eq:234}) can be treated perturbatively. The
results of our computations then will confirm this assumption.  Let us start
with eq.~(\ref{eq:233}).  With exception of the last term we have already
encountered similar system in Section~\ref{sec:final-state} (see also
Appendix~\ref{sec:final-state-compute}). For $|x\pm d|\gg \frac1{\m_\z}$ the
solution is given by analogs of
eqs.~(\ref{eq:51})--(\ref{eq:52}):\footnote{
  The result~(\ref{eq:239}) is easy to obtain, knowing the spectrum of the
  operator~(\ref{eq:169}) and writing its Green's function
  (compare~(\ref{eq:175})).}
\begin{equation}
\label{eq:239}
  \Phi_\gamma(x,z) = \phi_\gamma(x)
  \chi_0(z)+\kappa_\ggz\kappa_\gzz\chi_0(z)\int\frac{dp}{2\pi} 
  \frac{e^{i p x}\e^2\bH_0^2}{p^2 + m_0^2}\int\hs d\zeta\,
  \frac{\tilde\Phi_\z(p,\zeta)\chi_0(\zeta)}{\Delta(\zeta)\chi_0^2(0)} 
\end{equation}
(where by $\tilde \Phi_z(p,\zeta)$ we have denoted Fourier transform of the
function $\Phi_\z(x,\zeta)$ with respect to $x$). Function $\phi_\gamma(x)$
satisfies the Poisson equation with the mass term $\nabla^2 \phi_\gamma(x) -
m_0^2 \phi_\gamma(x)= q(x)$ with the mass
\begin{equation}
  \label{eq:240}
  m_0^2 = 12\kappa_\ggz e^2 \bH_0\:.
\end{equation}
The profile of $\Phi_\gamma$ in the $z$ direction
$\chi_0(z)$ is given by
\begin{equation}
\label{eq:241}
  \chi_0(z) = \frac{m_0^2 }{2\pi\M\sqrt{\Delta(z)}} 
  K_{\nu}( \tfrac{m_0^2}{4\M^2} e^{2\M|z|})K_{\nu}(\tfrac{m_0^2}{4\M^2}),
  \quad\text{where}\quad \nu = \frac12\sqrt{1-\frac{m_0^2}{\M^2}} 
\end{equation}
(here $K_\nu(u)$ is a modified Bessel function of the second kind). This
profile is sharply localized in the region
\begin{equation}
\label{eq:242}
  |z| \lesssim 
  \frac1{\M}\log\frac {2\M}{m_0}\:.
\end{equation}
The last term in~(\ref{eq:239}) 
can be treated perturbatively in $\kappa$ if a $z$-profile of the field
$\Phi_\z$ is decaying faster than $\Delta(z)$. To demonstrate that this is
indeed the case, let us insert the solution~(\ref{eq:239}) into the Gauss
constraint~(\ref{eq:228}). We will find that expression for $\Phi_\z$  is
given by expression, similar to the~(\ref{eq:239})--(\ref{eq:241}):
\begin{equation}
  \label{eq:243}
  \Phi_\z(x,z) =
  \phi_\z(x)\chi_{\z}(z)+\kappa_\ggz\kappa_\gzz\chi_{\z}(z)\int\frac{dp}{2\pi}
    \frac{e^{i p x}\e^2\bH_0^2}{p^2 + m_{0,\z}^2}\int\hs d\zeta\,
  \frac{\tilde\Phi_\gamma(p,\zeta)\chi_{\z}(\zeta)}{\Delta(\zeta)\chi_{\z}^2(0)}
\end{equation}
(by $\tilde\Phi_\gamma(p,\zeta)$ we have denoted Fourier transform of the
solution~(\ref{eq:239}) with respect to $x$).  Here $\chi_\z$ is a solution of
a differential equation similar to~(\ref{eq:167}) but with the modification of
the factor in front of the $\delta(z)$: $2\M\to 2 M- \m_\z$. This modifies
gluing condition~(\ref{eq:172}), which determined the eigen-values of the
equation in $z$ direction: $2\xi_0
K'_\nu(\xi_0)+\left(1-\frac{\m_\z}{2\M}\right)K_\nu(\xi_0)=0$. As a result
expression for $m_{0,\z}$ is given by:
\begin{equation}
  \label{eq:106}
  m_{0,\z}^2= \m_\z \M + 12 \kappa_\z e^2 \bH_0
\end{equation}
(the corrections to this result are of the order $\CO(\frac{\m_\z}{\M})$ and
$\CO(\kappa_\z\log\kappa_\z)$, where $\kappa_\z\equiv\sqrt{\kappa_\gzz^2 +
  \kappa_\ggz^2}$). As the first term in eq.~(\ref{eq:106}) is much bigger
than the second one\footnote{One can easily see this even for the smallest
  $\M\sim 10^4$~GeV and arbitrary $\kappa_0\lesssim 1$, recalling that
  1~Gauss$=2\cdot10^{-20}\:\mathrm{GeV}^2$ and the strongest magnetic fields,
  achievable in a laboratory are $\sim
  10^5$~Gauss.} 
we neglect the latter and find the mass $m_{0,\z}\gg \m_\z$ in contrast
with~(\ref{eq:240}). Notice, however, that the profile of the $\Phi_\z$ in $z$
direction is very similar to that of eq.~(\ref{eq:241}), still decaying as
$e^{-\Delta(z)}$ at the distances of the order~(\ref{eq:242}).  Finally,
function $\phi_\z$, entering~(\ref{eq:243}) satisfies Poisson equation with
the mass term of the \textsc{Z}-boson:
\begin{equation}
  \label{eq:244}
  \nabla^2 \phi_\z(x) - (\m_\z\M) \phi_\z(x)= q_\z(x)\:.
\end{equation}
%
Now we can analyze corrections due to the last term in eq.~(\ref{eq:243}).
From solution~(\ref{eq:239}) we see that both terms (and thus the whole
$\Phi_\gamma(x,z)$) have $z$-profile proportional to $\chi_0(z)$. Similarly,
$z$-profile of the field $\Phi_\z$ is proportional to $\chi_\z(z)$ and again
decays as $e^{-\Delta(z)}$ for $|z|$ outside the region~(\ref{eq:242}). One can
easily find that integrals over $\zeta$ in both expressions~(\ref{eq:239})
and~(\ref{eq:243}) behave as $\CO(\log\kappa_0)$\footnote{To estimate such
  integrals, it is enough to consider $\nu=\frac12$ as an index of a Bessel
  function in expressions for $\chi_0$ and $\chi_\z$.} 
and therefore these terms give small (of the order
$\CO(\kappa_0^2\log\kappa_0)$ corrections to the functions $\phi_\gamma(x)$
and $\phi_\z(x)$. Substituting expressions for $\Phi_\gamma$ and $\Phi_\z$
into (\ref{eq:225}) and~(\ref{eq:232}), one can easily see that these terms
are indeed proportional to $\kappa_0$, have the profiles, decaying as
$e^{-\Delta(z)}$ i.e. can be treated perturbatively.

Thus, we have shown that the last terms in eqs.~(\ref{eq:227})
and~(\ref{eq:228}) can be treated perturbatively in $\kappa_0$ and therefore
can be neglected. After that equation for $\Phi_\gamma$ becomes identical to
that of the Section~\ref{sec:final-state} and thus the static configuration of
an electric field in a capacitor will be the same!


\begin{thebibliography}{99}
%
\bibitem{YM} C.~N.~Yang and R.~L.~Mills, ``Conservation Of Isotopic Spin And
  Isotopic Gauge Invariance,'' Phys.\ Rev.\ {\bf 96}, 191 (1954).
  
\bibitem{gross-jackiw} D.~J.~Gross and R.~Jackiw, ``Effect Of Anomalies On
  Quasirenormalizable Theories,'' Phys.\ Rev.\ D {\bf 6}, 477 (1972);
%
\bibitem{abj}S.~L.~Adler, ``Axial Vector Vertex In Spinor Electrodynamics,''
  Phys.\ Rev.\ {\bf 177}, 2426 (1969);
%
  J.~S.~Bell and R.~Jackiw, ``A Pcac Puzzle: Pi0 $\to$ Gamma Gamma In The
  Sigma Model,'' Nuovo Cim.\ A {\bf 60}, 47 (1969).
%
%
\bibitem{thooft} G.~'t Hooft, ``Symmetry Breaking Through Bell-Jackiw
  Anomalies,'' Phys.\ Rev.\ Lett.\ {\bf 37}, 8 (1976).
  
\bibitem{rubakov} V.~A.~Rubakov, ``Monopole Induced Baryon Number
  Nonconservation,'' IYAI preprint IYAI-P-0211, April 1981;
  ``Adler-Bell-Jackiw Anomaly And Fermion Number Breaking In The Presence Of A
  Magnetic Monopole,'' Nucl.\ Phys.\ B {\bf 203}, 311 (1982).

\bibitem{callan} C.~G.~Callan, ``Disappearing Dyons,'' Phys.\ Rev.\ D {\bf
    25}, 2141 (1982);
  ``Dyon - Fermion Dynamics,'' \textit{ibid.}  {\bf 26}, 2058 (1982).

  
\bibitem{sm-anom}   
  C.~Bouchiat, J.~Iliopoulos and P.~Meyer, ``An Anomaly Free Version Of
  Weinberg's Model,'' Phys.\ Lett.\ B {\bf 38}, 519 (1972).

  
  H.~Georgi and S.~L.~Glashow, ``Gauge Theories Without Anomalies,'' Phys.\ 
  Rev.\ D {\bf 6}, 429 (1972).
  
\bibitem{df} E.~D'Hoker and E.~Farhi, ``Decoupling A Fermion In The Standard
  Electroweak Theory,'' Nucl.\ Phys.\ B {\bf 248}, 77 (1984);

E.~D'Hoker and E.~Farhi, ``Decoupling A Fermion Whose
  Mass Is Generated By A Yukawa Coupling: The General Case,'' Nucl.\ Phys.\ B
  {\bf 248}, 59 (1984).

  
\bibitem{rub-shap} V.~A.~Rubakov and M.~E.~Shaposhnikov, ``Do We Live
  Inside A Domain Wall?,'' Phys.\ Lett.\ B {\bf 125}, 136 (1983).
  
\bibitem{akama} K.~Akama, ``An Early Proposal Of 'Brane World','' Lect.\ Notes
  Phys.\ {\bf 176}, 267 (1982) [arXiv:hep-th/0001113].
  
\bibitem{Shap.qed} S.~Randjbar-Daemi and M.~Shaposhnikov, ``QED from
  six-dimensional vortex and gauge anomalies,'' JHEP {\bf 0304} (2003) 016
  [arXiv:hep-th/0303247].
  
\bibitem{fs} L.~D.~Faddeev and S.~L.~Shatashvili, ``Algebraic And Hamiltonian
  Methods In The Theory Of Nonabelian Anomalies,'' Theor.\ Math.\ Phys.\ {\bf
    60}, 770 (1984) [Teor.\ Mat.\ Fiz.\ {\bf 60}, 206 (1984)].

\bibitem{ch} C.~G.~Callan and J.~A.~Harvey, ``Anomalies And Fermion Zero Modes
  On Strings And Domain Walls,'' Nucl.\ Phys.\ B {\bf 250}, 427 (1985).

  
\bibitem{wen} X.~G.~Wen, ``Chiral Luttinger Liquid And The Edge Excitations In
  The Fractional Quantum Hall States,'' Phys.\ Rev.\ B {\bf 41}, 12838 (1990).
  
\bibitem{frohlich} J.~Frohlich and A.~Zee, ``Large scale physics of the
  quantum Hall fluid,'' Nucl.\ Phys.\ B {\bf 364}, 517 (1991).
  

\bibitem{girvin} R. E. Prange, S. M. Girvin, eds., {\it The Quantum Hall
    Effect}, (Springer, New York, 1990)
%
  
\bibitem{naculich} S.~Naculich, ``Axionic Strings: Covariant Anomalies And
  Bosonization Of Chiral Zero Modes,'' Nucl.\ Phys.\ B {\bf 296}, 837 (1988);
    
\bibitem{hr}
J.~A.~Harvey and O.~Ruchayskiy, ``The local structure of anomaly inflow,''
JHEP {\bf 0106}, 044 (2001) [arXiv:hep-th/0007037].

\bibitem{bhr}A.~Boyarsky, J.~A.~Harvey and O.~Ruchayskiy, ``A toy model of the
  M5-brane: Anomalies of monopole strings in five dimensions,'' Annals Phys.\ 
  {\bf 301}, 1 (2002) [arXiv:hep-th/0203154].

  
\bibitem{witten} E.~Witten,``Five-brane effective action in M-theory,'' J.\ 
  Geom.\ Phys.\ {\bf 22}, 103 (1997) [hep-th/9610234].

\bibitem{I-brane} M.~B.~Green, J.~A.~Harvey and G.~W.~Moore, ``I-brane
  inflow and anomalous couplings on D-branes,'' Class.\ Quant.\ Grav.\ {\bf
    14}, 47 (1997) [arXiv:hep-th/9605033].
  
\bibitem{fhmm} D.~Freed, J.~A.~Harvey, R.~Minasian and G.~Moore,
  ``Gravitational anomaly cancellation for M-theory fivebranes,'' Adv.\ 
  Theor.\ Math.\ Phys.\ {\bf 2}, 601 (1998) [hep-th/9803205].


  
\bibitem{bk} A.~Boyarsky and B.~Kulik, ``A note on the M5 brane anomaly,''
  Phys.\ Lett.\ B {\bf 516}, 171 (2001) [arXiv:hep-th/0107041].
  
\bibitem{anomaly-exp} A.~Boyarsky, O.~Ruchayskiy and M.~Shaposhnikov,
  ``Anomalies as a signature of extra dimensions,'' Phys.\ Lett.\ B {\bf 626},
  184 (2005) [arXiv:hep-ph/0507195].
  
\bibitem{theta-vacuum} S.~Khlebnikov and M.~Shaposhnikov, ``Brane-worlds and
  theta-vacua,'' Phys.\ Rev.\ D {\bf 71}, 104024 (2005)
  [arXiv:hep-th/0412306].


\bibitem{rubakov-rev} V.~A.~Rubakov, ``Large and infinite extra dimensions: An
  introduction,'' Phys.\ Usp.\ {\bf 44}, 871 (2001) [Usp.\ Fiz.\ Nauk {\bf
    171}, 913 (2001)] [arXiv:hep-ph/0104152].

\bibitem{r-ds} S.~Randjbar-Daemi and M.~Shaposhnikov, ``A
  formalism to analyze the spectrum of brane world scenarios,'' Nucl.\ Phys.\ 
  B {\bf 645}, 188 (2002) [arXiv:hep-th/0206016].


\bibitem{dubovsky}
S.~L.~Dubovsky, V.~A.~Rubakov and P.~G.~Tinyakov,
``Is the electric charge conserved in brane world?,''
JHEP {\bf 0008}, 041 (2000)
[arXiv:hep-ph/0007179].
  
\bibitem{charge-diff} M.~Marinelli and G.~Morpurgo,
``The Electric Neutrality Of Matter: A Summary,''
Phys.\ Lett.\ B {\bf 137}, 439 (1984).
  

\bibitem{oda} I.~Oda, ``Localization of matters on a string-like defect,''
  Phys.\ Lett.\ B {\bf 496}, 113 (2000) [arXiv:hep-th/0006203].

\bibitem{st} M.~E.~Shaposhnikov and P.~Tinyakov, ``Extra dimensions as an
  alternative to Higgs mechanism?,'' Phys.\ Lett.\ B {\bf 515}, 442 (2001)
  [arXiv:hep-th/0102161].

  
\bibitem{laine} M.~Laine, H.~B.~Meyer, K.~Rummukainen and
  M.~Shaposhnikov, ``Localisation and mass generation for non-Abelian gauge
  fields,'' JHEP {\bf 0301}, 068 (2003) [arXiv:hep-ph/0211149].

\bibitem{dvali-localization} G.~R.~Dvali, G.~Gabadadze and M.~A.~Shifman,
  ``(Quasi)localized gauge field on a brane: Dissipating cosmic radiation to
  extra dimensions?,'' Phys.\ Lett.\ B {\bf 497}, 271 (2001)
  [arXiv:hep-th/0010071].

  
\bibitem{redlich} A.~N.~Redlich, ``Gauge Noninvariance And Parity
  Nonconservation Of Three-Dimensional Fermions,'' Phys.\ Rev.\ Lett.\ {\bf
    52}, 18 (1984);
  
  A.~N.~Redlich, ``Parity Violation And Gauge Noninvariance Of The Effective
  Gauge Field Action In Three-Dimensions,'' Phys.\ Rev.\ D {\bf 29} (1984)
  2366.
  
\bibitem{a-gd-pm}L.~Alvarez-Gaume, S.~Della Pietra and G.~W.~Moore,
  ``Anomalies And Odd Dimensions,'' Annals Phys.\ {\bf 163}, 288 (1985).


\bibitem{cov-anom}W.~A.~Bardeen and B.~Zumino, ``Consistent And Covariant
  Anomalies In Gauge And Gravitational Theories,'' Nucl.\ Phys.\ B {\bf 244},
  421 (1984).
  
\bibitem{amb-Higgs} J.~Ambjorn and P.~Olesen, ``Electroweak Magnetism: Theory
  And Application,'' Int.\ J.\ Mod.\ Phys.\ A {\bf 5}, 4525 (1990).


\bibitem{amb-anomaly} J.~Ambjorn, J.~Greensite and C.~Peterson, ``The Axial
  Anomaly And The Lattice Dirac Sea,'' Nucl.\ Phys.\ B {\bf 221}, 381 (1983).
  
\bibitem{gradshteyn} I.~S.~Gradshteyn, I.~M.~Ryzhik, A.~Jeffrey,
  D.~Zwillinger, \emph{Table of Integrals, Series, and Products}, any edition.

  
\bibitem{gauge-fields} A.~Boyarsky, O.~Ruchayskiy, M.~Shaposhnikov, ``Gauge
  fields in brane-worlds'', \emph{to appear}
  
\bibitem{Appelquist} T.~Appelquist and J.~Carazzone, ``Infrared Singularities
  And Massive Fields,'' Phys.\ Rev.\ D {\bf 11}, 2856 (1975).

\bibitem{Antoniadis} I.~Antoniadis and S.~Dimopoulos, ``Splitting
  supersymmetry in string theory,'' arXiv:hep-th/0411032.
  
\end{thebibliography}
\end{document}